\documentclass[11pt,a4paper]{article}
\pdfoutput=1

\usepackage[utf8]{inputenc}
\usepackage[T1]{fontenc}
\usepackage{lmodern}

\usepackage{jheppub}
\usepackage{graphicx,amssymb,amsmath,amsfonts}
\usepackage{amsthm}
\usepackage[greek, english]{babel}
\usepackage{hyperref}
\usepackage{cleveref}
\usepackage{subcaption}

\usepackage{xfrac}
\usepackage{braket}
\usepackage{xcolor}

\newcommand{\mn}{{\mu\nu}}
\newcommand{\rs}{{\rho\sigma}}
\newcommand{\ab}{{\alpha\beta}}
\newcommand{\mnrs}{{\mu\nu\rho\sigma}}

\newcommand{\V}{\mathbf}

\newcommand{\p}{{\partial}}

\newcommand{\ep}{\epsilon}

\newcommand{\zb}{{\bar{z}}}
\newcommand{\wb}{{\bar{w}}}
\newcommand{\gzz}{{\gamma_{z\bar{z}}}}
\newcommand{\guzz}{{\gamma^{z\bar{z}}}}

\newcommand{\mI}{\mathcal{I}}

\newcommand{\mL}{\mathcal{L}}
\newcommand{\mM}{\mathcal{M}}
\newcommand{\mN}{\mathcal{N}}
\newcommand{\mO}{O}

\newcommand{\mQ}{\mathcal{Q}}

\newcommand{\mS}{\mathcal{S}}

\newcommand{\w}{{\omega}}
\newcommand{\hc}{\text{h.c.}}
\newcommand{\cc}{\text{c.c.}}
\newcommand{\zz}{{(z\to\bar{z})}}

\newcommand{\tin}{{\text{in}}}
\newcommand{\tout}{{\text{out}}}

\newcommand{\td}[1]{\widetilde{d^3 #1}\,}

\newcommand{\llangle}{\langle\mkern-4mu\langle}
\newcommand{\rrangle}{\rangle\mkern-4mu\rangle}
\newcommand{\bbraket}[1]{{\left\llangle #1 \right\rrangle}}
\newcommand{\kket}[1]{{\left\Vert #1\right\rrangle}}
\newcommand{\bbra}[1]{{\left\llangle #1 \right\Vert}}

\newcommand{\Esoft}{{\Lambda}}
\newcommand{\cutoff}{{\lambda}}
\newcommand{\Eres}{{E_\text{res}}}
\newcommand{\Yg}{Y_g}
\newcommand{\Ye}{Y_e}

\title{Subleading soft dressings of asymptotic states in QED and perturbative quantum gravity}
\author{Sangmin Choi}
\author{and Ratindranath Akhoury}

\affiliation{Leinweber Center for Theoretical Physics, \\
Randall Laboratory of Physics, Department of Physics,\\
University of Michigan, Ann Arbor, MI 48109, USA}

\emailAdd{sangminc@umich.edu}
\emailAdd{akhoury@umich.edu}

\abstract{
	We construct Faddeev-Kulish states in QED and perturbative quantum gravity to subleading order in the soft momentum expansion
		and to first order in the coupling constant,
		using the charge conservation formula of asymptotic symmetries associated with the tree-level subleading soft theorems.
	We demonstrate that the emission and absorption of soft photons/gravitons in dressed amplitudes vanish.
	The fact that no additional soft radiation may be added to a dressed amplitude
		supports the claim that, in the dressed state formalism, the soft and hard sectors of scattering processes are correlated.
	We also show that the dressed virtual amplitudes are equivalent to the infrared-finite part of the traditional amplitudes constructed using Fock states.
	Since there is no real soft radiation in the asymptotic Hilbert space, the dressed state formalism gives the same cross sections as
		the Bloch-Nordsieck method.
}

\begin{document}
{
	\maketitle
}

\section{Introduction}\label{sec:intro}

It is well known that in the conventional formulation of quantum field theories, S-matrix elements are plagued by infrared divergences
	and are therefore ill-defined.
This is due to the fact that interactions in such theories cannot be turned off at large times,
	which invalidates the use of Fock states as asymptotic states.
In 1970, Kulish and Faddeev \cite{Kulish:1970ut} showed for QED that the correct asymptotic Hilbert space which accounts for
	the non-vanishing interactions is the set of ``dressed'' states, which are essentially coherent states of soft photons.
In 2013, an analogous construction has been made for perturbative quantum gravity by \cite{Ware:2013zja}.
Such coherent states are referred to as Faddeev-Kulish (FK) states, or dressed states.
Indeed, it was shown by Chung \cite{Chung:1965zza} for QED and Ware et al$.$ \cite{Ware:2013zja} for gravity that
	scattering amplitudes in the FK state basis do not exhibit infrared divergence.
Throughout the paper, we will refer to such amplitudes as FK amplitudes, as opposed to the traditional amplitudes constructed using Fock states.

The FK states are closely related to asymptotic symmetries of gauge theories.
In recent years, asymptotic symmetries have received a considerable amount of attention
	due to their subtle connection with soft theorems
	\cite{Strominger:2013jfa,He:2014laa,He:2014cra,Kapec:2014opa,Strominger:2013lka,Kapec:2015ena,He:2017fsb,
	Cachazo:2014fwa,Lysov:2014csa,Strominger:2014pwa,Kapec:2014zla,Kapec:2015vwa,He:2015zea,Strominger:2015bla,Dumitrescu:2015fej,
	Bianchi:2014gla,
	Campiglia:2014yka,Campiglia:2015kxa,Campiglia:2015qka,Campiglia:2015yka,Campiglia:2016jdj,Laddha:2017vfh,Campiglia:2017mua,Campiglia:2017dpg,
	Campiglia:2017xkp,Ashtekar:2018lor,Campiglia:2018see,Campiglia:2019wxe,
	He:2019jjk,He:2019pll}.
The soft charges associated with the asymptotic symmetries induce a degeneracy in the vacuum state of the theory,
	and charge conservation leads to a transition among the vacua in a scattering process.
It has been shown that the FK states implement this vacuum transition properly, while Fock states do not
	\cite{Gabai:2016kuf,Kapec:2017tkm,Choi:2017bna,Choi:2017ylo}; FK dressings are operators that carry a definite leading soft charge.
In this context, vanishing S-matrix elements in Fock state basis due to infrared divergence can be regarded as
	a consequence of a superselection rule of the degenerate vacua.
This relationship between asymptotic symmetries and FK dressings have been shown to be generalizable to other asymptotic boundaries of the
	spacetime, such as Rindler and black hole horizons \cite{Blommaert:2018oue,Blommaert:2018rsf,Choi:2018oel}.

Although the FK state approach has proven to be useful, there are still unanswered questions.
For example, in the traditional construction of amplitudes, one deals with infrared divergences by summing over
	inclusive cross sections \cite{BN}, leaving only the contribution from the infrared-finite part of the amplitude \cite{Weinberg:1965nx}.
The FK amplitude is free of infrared divergence, but does it agree with this infrared-finite part?
For another example, given a set of dressed states as asymptotic states, may one add arbitrary soft radiation to the incoming state?
Also, do the soft particles carry information about the hard particles?
To answer these questions, the known FK states, which only contain terms at the leading order in the soft expansion, are not enough;
	one needs to extend the construction to the subleading order.

Motivated by these questions, in this paper we derive the FK dressings of QED and gravity to
	first order in the coupling and subleading order in the soft expansion.
The known leading soft FK states have been shown to be
	charge eigenstates of asymptotic symmetry associated with the leading soft theorem \cite{Kapec:2017tkm,Choi:2017ylo}.
Therefore, one would expect that a similar construction using asymptotic symmetries associated with the subleading soft theorem
	will correctly lead to the subleading dressing.
However, this presents several challenges.
The subleading soft charges do not commute with each other as well as with the leading soft charges,
	so one cannot construct simultaneous eigenstates that carry a definite leading and subleading charge.
But as discussed in section \ref{sec:gravity}, to first order in the coupling we may finesse this problem as far as S-matrix elements
	are concerned, and show that the construction is self consistent.
The subleading soft dressing operator so obtained agrees with the Wilson line dressing first derived
	by Mandelstam \cite{Mandelstam:1962us} to first order in the coupling in the exponent.
Thus to this order in the coupling and in the soft expansion, there is again the equivalence between Wilson line dressings
	and FK dressings discussed in \cite{Jakob:1990zi,Choi:2018oel}.
With the subleading dressings at hand, we proceed to derive two main results, which we outline in the next two paragraphs.

First, we show that an external photon/graviton insertion to an FK amplitude vanishes in the soft limit.
It follows that once Fock states are dressed to yield a set of FK states,
	there is no extra soft radiation in the asymptotic Hilbert space.
That is, any soft photons/gravitons may only appear through FK dressings.
We anticipate that this result may shed some light on the discussions regarding factorization of the soft and hard sectors,
	and whether the soft particles carry information of the hard degrees of freedom;
	see \cite{Mirbabayi:2016axw,Bousso:2017dny,Carney:2017jut,Carney:2017oxp} for example.
We can say that the soft and hard sectors are correlated for the FK states since specifying one fixes the other.
Also, since a soft photon/graviton creation operator acting on a vacuum makes it orthogonal to all states in the asymptotic Hilbert space,
	one may assume that it annihilates the vacuum as far as scattering processes are concerned.
This removes some complications regarding Fock vacuum and soft operators, see \cite{Gabai:2016kuf} for example.

Second, we show that the virtual FK amplitude is equivalent to the infrared-finite part of the traditional amplitude in the Fock basis,
	up to power-law type corrections of the soft energy scale which is negligible by definition.
This is rather reassuring; if this were not the case, FK amplitudes would be in conflict with
	experimentally measurable cross sections, which involve the infrared-finite part of traditional amplitudes.
Since there is no real soft radiation in the asymptotic Hilbert space, the FK state formalism yields the same cross sections
	as the Bloch-Nordsieck method \cite{BN}.
This serves as a sanity check for the argument that the asymptotic states are FK states instead of Fock states.

Two important comments are in order.
First, it has been shown that the subleading soft theorems receive one-loop exact quantum corrections \cite{Bern:2014oka}
	that are logarithmically divergent in the soft energy.
Since we construct the subleading dressing only to first order in the coupling constant,
	we will not consider such corrections and work with the tree-level subleading soft theorems.
While we leave details of the loop-corrected dressings for future investigation,
	we argue that our second result, that FK amplitudes are in agreement with infrared-finite part of traditional amplitudes,
	should remain unaffected by loop corrections.
This follows from the fact that logarithmic divergence is an integrable singularity,
	and consequently the additional contributions of loop corrections to FK amplitudes integrate to negligible quantities.
Secondly, there is a piece of the subleading soft contributions to scattering amplitudes which does not exponentiate.
This is just the contribution usually assigned to Low's theorem \cite{Low1,Low2,GellMannGoldberger}.
This complication, again, may be ignored at leading order in the coupling constant in the exponent of the dressing operator.

The paper is organized as follows.
In section \ref{sec:gravity} we begin with a brief review of superrotation and the associated charge,
	followed by the construction of the gravitational subleading FK dressing.
Section \ref{sec:QED} presents an analogous construction for the subleading FK dressing in QED.
Using the dressings, in section \ref{sec:external} we show that the insertion of a soft graviton makes the amplitude vanish,
	and in section \ref{sec:equiv} we demonstrate the equivalence between FK amplitudes and the finite part of traditional amplitudes.
The dressings of gravity and QED are very similar; although sections \ref{sec:external} and \ref{sec:equiv} work explicitly with gravitons,
	we expect the conclusions to be valid for QED as well.
In appendix \ref{app:ssfnsam} we review the correspondence between subleading soft theorems and the Ward identities of the associated asymptotic symmetries,
	both for scalars and for particles with spin.

\section{Dressings in perturbative quantum gravity}\label{sec:gravity}

\subsection{Review of superrotation in asymptotically flat spacetime}

We start by establishing our notation regarding asymptotically flat spacetimes
	and reviewing the materials associated with the superrotation on $\mI^\pm$.
We follow the construction of \cite{Kapec:2014opa} closely.
For the sake of simplicity we will assume that all matter particles are massless scalars; for particles with spin
	see appendix \ref{app:gravspin}.

\subsubsection{Metric and mode expansions}

In Bondi coordinates, the metric for an asymptotically flat spacetime near the future null infinity $\mI^+$ reads \cite{Kapec:2014opa}
\begin{align}
	ds^2 &= -du^2 - 2dudr + 2r^2\gzz dzd\zb
		\nonumber\\&\quad
		+\frac{2m^+_B}{r}du^2 + rC_{zz}dz^2 + rC_{\zb\zb}d\zb^2 + 2g_{uz}dudz + 2g_{u\zb}dud\zb + \cdots,
\end{align}
where the first line corresponds to the flat metric.
Here $z=e^{i\phi}\tan\frac{\theta}{2}$ is the stereographic coordinate, $\gzz=\frac{2}{(1+z\zb)^2}$ is the metric on the 2-sphere
	and $m^+_B$ is the Bondi mass aspect.
The $uz$-component of the metric has the expansion
\begin{align}
	g_{uz} = \frac{1}{2}D^zC_{zz} + \frac{1}{6r}C_{zz}D_zC^{zz} + \frac{2}{3r}N^+_z + \cdots,
\end{align}
where $D_z$ is the covariant derivative on $S^2$ and $N^+_z$ is the angular momentum aspect.
In general, $m^+_B$, $N^+_z$ and $C_{zz}$ are functions of $u$, $z$ and $\zb$.

Let us define the graviton field $h_\mn$ through
\begin{align}
	g_\mn(x) = \eta_\mn + \kappa h_\mn(x),\qquad\kappa^2=32\pi G,
\end{align}
where $\eta_\mn = \mathrm{diag}(-1,1,1,1)$.
Near $\mI^+$, the graviton field can be approximated by the on-shell mode expansion
\begin{align}
	h^\tout_\mn(x) = \sum_{s=\pm}\int \frac{d^3k}{(2\pi)^3}\frac{1}{2\w_k}
		\left[\ep_\mn^{s*}(\V k)a^\tout_s(\V k)e^{ik\cdot x} + \ep_\mn^s(\V k) a^{\tout\dagger}_s(\V k)e^{-ik\cdot x}\right],
\end{align}
where $\w_k = k^0 = |\V k|$, and $\ep^\pm_\mn=\ep^\pm_\mu\ep^\pm_\nu$ are the spin-2 polarization tensors.
By parametrizing the graviton momentum $k^\mu$ by $(\w_k,z,\zb)$,
\begin{align}
	k^\mu = \frac{\w_k}{1+z\zb}\Big(1+z\zb,\zb+z,i(\zb-z),1-z\zb\Big),
\end{align}
we may write the polarization tensors as
\begin{align}
	\ep^{+\mu}(\V k) = \frac{1}{\sqrt 2}(\zb,1,-i,-\zb),\qquad
	\ep^{-\mu}(\V k) = \frac{1}{\sqrt 2}(z,1,i,-z).
\end{align}
The out-operators satisfy the standard commutation relation,
\begin{align}
	\left[a^\tout_s(\V k),a^{\tout\dagger}_r(\V k')\right] = \delta_{sr}(2\pi)^3(2\w_k)\delta^{(3)}(\V k-\V k').
\end{align}

Near the past null infinity $\mI^-$, the asymptotically flat metric reads
\begin{align}
	ds^2 &= -dv^2 + 2dvdr + 2r^2\gzz dzd\zb
		\nonumber\\&\quad
		+\frac{2m^-_B}{r}dv^2 + rD_{zz}dz^2 + rD_{\zb\zb}d\zb^2 + 2g_{vz}dvdz + 2g_{v\zb}dvd\zb + \cdots,
\end{align}
where
\begin{align}
	g_{vz} = -\frac{1}{2}D^z D_{zz} - \frac{1}{6r}D_{zz}D_zD^{zz} - \frac{2}{3r}N_z^- + \cdots.
\end{align}
We have the mode expansion for the incoming graviton field
\begin{align}
	h^\tin_\mn(x) = \sum_{s=\pm}\int \frac{d^3k}{(2\pi)^3}\frac{1}{2\w_k}
		\left[\ep_\mn^{s*}(\V k)a^\tin_s(\V k)e^{ik\cdot x} + \ep_\mn^s(\V k) a^{\tin\dagger}_s(\V k)e^{-ik\cdot x}\right],
\end{align}
where the in-operators satisfy
\begin{align}
	\left[a^\tin_s(\V k),a^{\tin\dagger}_r(\V k')\right] = \delta_{sr}(2\pi)^3(2\w_k)\delta^{(3)}(\V k-\V k').
\end{align}

\subsubsection{Superrotation charge}

Superrotation near $\mI^\pm$ are generated by the following vector fields respectively \cite{Kapec:2014opa},
\begin{align}
	\xi^+(Y) &=
		\left(1+\frac{u}{2r}\right)Y^z\p_z
		- \frac{u}{2r}D^\zb D_z Y^z\p_\zb
		- \frac{(r+u)}{2}D_z Y^z\p_r + \frac{u}{2}D_z Y^z\p_u
		+ \cc,
	\\
	\xi^-(Y) &=
		\left(1-\frac{v}{2r}\right)Y^z\p_z + \frac{v}{2r}D^\zb D_zY^z\p_\zb - \frac{(r-v)}{2}D_zY^z\p_r + \frac{v}{2}D_zY^z\p_v + \cc,
\end{align}
parametrized by the same vector $Y^z$ on the sphere.
We will drop the constraint that $Y^z$ is a conformal Killing vector, following \cite{Campiglia:2014yka}.
The conserved charges associated with superrotation have the expressions \cite{Strominger:2017zoo}
\begin{align}
	Q^\pm_Y &= \frac{4}{\kappa^2}\int_{\mI^\pm_\mp}d^2z\left(Y_\zb N^\pm_z + Y_z N^\pm_\zb\right),
\end{align}
where $\mI^+_-$ ($\mI^-_+)$ is the past (future) boundary of the future (past) null infinity.
The charges can be decomposed into soft and hard parts,
\begin{align}
	Q_Y^\pm &= Q_S^\pm + Q_H^\pm,
	\label{Q_Y}
\end{align}
where the soft charges are given by
\begin{align}
	Q_S^+ &= -\frac{2}{\kappa^2} \int_{\mI^+} du\,d^2z\, \guzz D_z^3Y^z u N_{\zb\zb} + \hc
		\\ &=
			-\frac{i}{4\pi\kappa}
			\lim_{\w\to 0}(1+\w\p_\w)
			\int d^2z\, D_z^3Y^z
			\left[a^\tout_-(\w\V x_z) - a^{\tout\dagger}_+(\w\V x_z)\right] + \hc,
	\label{soft}
	\\
	Q_S^- &= \frac{2}{\kappa^2} \int_{\mI^-} dv\,d^2z\, \guzz D_z^3Y^z v M_{\zb\zb} + \hc
	\\ &=
		\frac{i}{4\pi\kappa}
		\lim_{\w\to 0}(1+\w\p_\w)
		\int d^2z\, D_z^3Y^z
		\left[a^\tin_-(\w\V x_z) - a^{\tin\dagger}_+(\w\V x_z)\right] + \hc,
	\label{soft2}
\end{align}
with $N_{zz}=\p_u C_{zz}$ and $M_{zz}=\p_v D_{zz}$; we refer to \cite{Kapec:2014opa} for details.
Here $\V x_z$ denotes a unit 3-vector whose direction is given by $(z,\zb)$,
\begin{align}
	\V x_z = \frac{1}{1+z\zb}\Big(\zb+z,i(\zb-z),1-z\zb\Big).
	\label{x_z}
\end{align}
The hard charges $Q_H^\pm$ are defined by their actions on the Fock states,
\begin{align}
	\bra{\V p_1,\ldots,\V p_n} Q_H^+ &=
		i\sum_{i=1}^n\left(Y^z(z_i)\p_{z_i}-\frac{E_i}{2}D_z Y^z(z_i)\p_{E_i}+\zz\right)
		\bra{\V p_1,\ldots,\V p_n},
	\label{hard}
	\\
	Q_H^- \ket{\V p_1,\ldots,\V p_n} &=
		-i\sum_{i=1}^n\left(Y^z(z_i)\p_{z_i}-\frac{E_i}{2}D_z Y^z(z_i)\p_{E_i}+\zz\right)
		\ket{\V p_1,\ldots,\V p_n}.
	\label{hard2}
\end{align}
where the momentum of the massless scalar $p_i$ is written as
\begin{align}
	p_i^\mu = \frac{E_i}{1+z_i\zb_i}\Big(1+z_i\zb_i,\zb_i+z_i,i(\zb_i-z_i),1-z_i\zb_i\Big).
\end{align}

\subsection{Distinction between in and out operators} \label{sec:distinction}

Notice that in \eqref{soft} and \eqref{soft2} we follow the notation of \cite{Strominger:2013jfa} and others
	to distinguish between the ``out'' operators on $\mI^+$ and the ``in'' operators on $\mI^-$.
They are related by the boundary condition $N_{\zb\zb}|_{\mI^+_-}=-M_{\zb\zb}|_{\mI^-_+}$ such that the subleading soft contribution to the S-matrix element
	from insertions of $a^{\tin\dagger}_s$ in the incoming state and $-a^{\tout}_{-s}$ in the outgoing state are equivalent,
	see \cite{Kapec:2014opa} for a discussion.
An alternative approach, which was taken in \cite{Choi:2017ylo}, is to make this relation explicit by taking in \eqref{soft2} and \eqref{soft3},
\begin{align}
	a^\tout_s(\w\V x_z) \to a_{s}(\w\V x_z),
	\qquad
	a^\tin_s(\w\V x_z) \to -a_{s}(\w\V x_z),
	\label{conv}
\end{align}
such that $Q_S^+=(Q_S^-)^\dagger=Q_S^-$, and
\begin{align}
	\left[a_r(\V k),a_s^\dagger(\V k')\right]=\delta_{rs}(2\pi)^3(2\w_k)\delta^{(3)}(\V k-\V k').
\end{align}
Then we can remove the superscript in $Q^\pm_S$ and write
\begin{align}
	Q_S &=
		\frac{-i}{4\pi\kappa}
		\lim_{\w\to 0}(1+\w\p_\w)
		\int d^2z\, D_z^3Y^z
		\left[a_-(\w\V x_z) - a^{\dagger}_+(\w\V x_z)\right] + \hc.
	\label{soft3}
\end{align}
Either approach will yield the same result;
	with the distinction intact, one just has to be cautious when contracting an ``out'' and an ``in'' operator.
We will employ the latter convention \eqref{conv} and use \eqref{soft3} for it removes some complications in the amplitude computation.

\subsection{Subleading soft dressing}\label{sec:grav_subleading}

Let us consider the scattering from some incoming state $\ket{\tin}$ to some outgoing state $\bra{\tout}$ -- the states can be either
	dressed states or Fock states.
Superrotation symmetry states that the associated charge must be conserved in a scattering process, i.e.
\begin{align}
	\braket{\tout|\left(Q^+_Y\mS - \mS Q_Y^-\right)|\tin}=0,
\end{align}
where $\mS$ is the scattering matrix.
This can be written as
\begin{align}
	\braket{\tout|[Q_S(Y),\mS]|\tin} &= -i\sum_i \left(Y^z(z_i)\p_{z_i}-\frac{E_i}{2}D_zY^z(z_i)\p_{E_i}+\zz\right)\braket{\tout|\mS|\tin},
	\label{charge_conservation}
\end{align}
where we used \eqref{hard} and \eqref{hard2}.

Let us choose the vector field
\begin{align}
	Y = \Yg \equiv \frac{(z-w)^2}{(\zb-\wb)}\p_z,
	\label{choice_grav}
\end{align}
for which \eqref{charge_conservation} becomes \cite{Kapec:2014opa,Campiglia:2014yka} (see appendix \ref{app:grav} for a derivation)
\begin{align}
	\braket{\tout|[Q_S(\Yg),\mS]|\tin} &=
		-\sum_i\frac{p_i^\mu k_\lambda\ep^-_\mn(\w\V x_z)}{p_i\cdot k}
		\left(
			p_i^\lambda\frac{\p}{\p p_{i\nu}}
			- p_i^\nu \frac{\p}{\p p_{i\lambda}}
		\right)\braket{\tout|\mS|\tin}
	\\ &=
		-iS_g^{(1)-}\braket{\tout|\mS|\tin},
		\label{charge_conservation2}
\end{align}
where $k^\mu\equiv(\w,\w\V x_z)$, and $S_g^{(1)-}$ is the subleading soft factor for negative-helicity graviton,
\begin{align}
	S_g^{(1)-}=-i\sum_i\eta_i\frac{p_i^\mu k_\lambda J_i^{\lambda\nu}}{p_i\cdot k}\ep^-_\mn(\w\V x_z),
	\label{subleading_softfactor}
\end{align}
with $\eta_i=+1$ for incoming particles and $\eta_i=-1$ for outgoing particles.\footnote{
	In this definition of $S_g^{(1)-}$ we deviate from the convention used in \cite{Kapec:2014opa}.
	The signs $\eta_i$ derive from the different momentum-space representations of the action of angular momentum on bras and kets.
	Alternatively, one could define $J_i^{\lambda \nu}$ as in \cite{Kapec:2014opa}, for which case the statement of angular momentum conservation
		becomes $\sum_i J_i=0$.
	We will adopt \eqref{subleading_softfactor}, in order for angular momentum conservation to take the more natural form
		$\sum_{i\in\tin}J_i=\sum_{j\in\tout}J_j$.
	The two approaches are equivalent.
	\label{eta_origin}
	}
Using the identity $D_z^3\Yg^z=4\pi\delta^{(2)}(z-w)$, we may write the soft charge as
\begin{align}
	Q_S(\Yg) &=
			-\frac{i}{\kappa}
			\lim_{\w\to 0}(1+\w\p_\w)
			\left[a_-(\w\V x_z) - a_+^\dagger(\w\V x_z)\right].
\end{align}

We now claim that under certain circumstances stated below, we may write for a ket vacuum $\ket{0}$,
\begin{align}
	Q_S\ket{0}\approx 0.
\end{align}
Strictly speaking, the subleading soft charge does not annihilate the vacuum state (and hence the symbol $\approx$),
	but rather creates a state containing a soft graviton.
In section \ref{sec:external} we will show that no state may scatter to such a state and vice versa
	in the dressed state formalism.
Therefore, $Q_S$ may be taken to annihilate $\ket{0}$ insofar as scattering processes are concerned.

As was done in \cite{Choi:2017ylo} for the leading soft dressings,
	we aim to construct the subleading soft dressed state by using superrotation charge conservation.
Since $Q_S(\Yg)$ is of the form $a - a^\dagger$, we want to consider a coherent state of the form
\begin{align}
		\exp\left\{
			\frac{i\kappa}{2}\sum_{s={\pm}}
			\int\frac{d^3k}{(2\pi)^3}\frac{\phi(\w_k)}{2\w_k}
			N^\mn_\tin \left[
				\epsilon^{s*}_\mn(\V k) a_s(\V k)
				+ \epsilon^s_\mn(\V k) a_s^\dagger(\V k)
			\right]
		\right\}\ket{\V p_1,\cdots,\V p_m},
	\label{exp_grav}
\end{align}
where $N^\mn_\tin$ is a tensor whose components are to be determined by charge conservation,
	$\phi(\w_k)$ is an infrared function \cite{Kulish:1970ut, Ware:2013zja} that has support only in a small neighborhood of $\w_k=0$
	satisfying $\phi(0)=1$.
Here $\ket{\V p_1,\cdots,\V p_m}$ is an $m$-particle Fock state,
\begin{align}
	\ket{\V p_1,\cdots,\V p_m} = \prod_{i=1}^m b^\dagger(\V p_i)\ket{0},
\end{align}
where $b^\dagger(\V p)$ is the creation operator of the scalar field.
However, since we are using the tree-level subleading soft theorem, we can only construct a dressing that can be trusted
	to order $\kappa$ in the exponent.
In this spirit, let us define the incoming state as the linearized version of \eqref{exp_grav},
\begin{align}
	\ket{\tin} &=
		\left\{
			1+
			\frac{i\kappa}{2}\sum_{s={\pm}}
			\int\frac{d^3k}{(2\pi)^3}\frac{\phi(\w_k)}{2\w_k}
			N^\mn_\tin \left[
				\epsilon^{s*}_\mn(\V k) a_s(\V k)
				+ \epsilon^s_\mn(\V k) a_s^\dagger(\V k)
			\right]
		\right\}
		\ket{\V p_1,\cdots,\V p_m}.
		\label{grav_linear}
\end{align}
By direct computation,
\begin{align}
	Q_S(\Yg)\ket{\tin} &=
			\frac{1}{2}
			\lim_{\w\to 0}(1+\w\p_\w)
			\sum_{s={\pm}}
			\int\frac{d^3k}{(2\pi)^3}\frac{\phi(\w_k)}{2\w_k}N^\mn_\tin
		\nonumber\\&\quad\times
			\left[
				a_-(\w\V x_z) - a_+^\dagger(\w\V x_z),
				\epsilon^{s*}_\mn(\V k) a_s(\V k) + \epsilon^s_\mn(\V k) a_s^\dagger(\V k)
			\right]
			\ket{\V p_1,\cdots,\V p_m}
		\\ &=
			\frac{1}{2}
			\lim_{\w\to 0}(1+\w\p_\w)
			\sum_{s={\pm}}
			\int\frac{d^3k}{(2\pi)^3}\frac{\phi(\w_k)}{2\w_k}N^\mn_\tin
		\nonumber\\&\quad\times
			(2\pi)^3(2\w_k)\delta^{(3)}(\V k-\w\V x_z)
			\left(
				\epsilon_\mn^s(\V k)\delta_{s,-}
				+ \epsilon_\mn^{s*}(\V k)\delta_{s,+}
			\right)
			\ket{\V p_1,\cdots,\V p_m}
		\\ &=
			\lim_{\w\to 0}(1+\w\p_\w)
			N_\tin\cdot\epsilon^-(\w\V x_z)
			\ket{\V p_1,\cdots,\V p_m},
\end{align}
where in the last line we used the notation $N_\tin\cdot \epsilon^-\equiv N_\tin^\mn \epsilon_\mn^-$.

Similarly, we can construct a bra state,
\begin{align}
	\bra{\tout} &\equiv
		\bra{\V p_{m+1},\cdots,\V p_{m+n}}
		\left(
			1
			-\frac{i\kappa}{2}\sum_{s={\pm}}
			\int\frac{d^3k}{(2\pi)^3}\frac{\phi(\w_k)}{2\w_k}
			N^\mn_\tout \left[
				\epsilon^{s*}_\mn(\V k) a_s(\V k)
				+ \epsilon^s_\mn(\V k) a_s^\dagger(\V k)
			\right]
		\right),
\end{align}
such that
\begin{align}
	\bra{\tout}Q_S(\Yg) &=
			-\frac{1}{2}
			\lim_{\w\to 0}(1+\w\p_\w)
			\sum_{s={\pm}}
			\int\frac{d^3k}{(2\pi)^3}\frac{\phi(\w_k)}{2\w_k}
			\bra{\V p_{m+1},\cdots,\V p_{m+n}}N^\mn_\tout
		\nonumber\\&\quad\times
			\left[
				\epsilon^{s*}_\mn(\V k) a_s(\V k) + \epsilon^s_\mn(\V k) a_s^\dagger(\V k),
				a_-(\w\V x_z) - a_+^\dagger(\w\V x_z)
			\right]
		\\ &=
			\frac{1}{2}
			\lim_{\w\to 0}(1+\w\p_\w)
			\sum_{s={\pm}}
			\int\frac{d^3k}{(2\pi)^3}\frac{\phi(\w_k)}{2\w_k}
			(2\pi)^3(2\w_k)\delta^{(3)}(\V k-\w\V x_z)
		\nonumber\\&\quad\times
			\left(
				\epsilon_\mn^{s*}(\V k)\delta_{s,+}
				+ \epsilon_\mn^s(\V k)\delta_{s,-}
			\right)
			\bra{\V p_{m+1},\cdots,\V p_{m+n}}N^\mn_\tout
		\\ &=
			\lim_{\w\to 0}(1+\w\p_\w)
			\bra{\V p_{m+1},\cdots,\V p_{m+n}}
			N_\tout\cdot\epsilon^-(\w\V x_z).
\end{align}
With these states, we may write
\begin{align}
	\bra{\tout}[Q_S(\Yg),\mS]\ket{\tin}
		&= \lim_{\w\to 0}(1+\w\p_\w)
			\Big[
				\braket{\V p_{m+1},\cdots,\V p_{m+n}|(N_\tout\cdot \ep^-)\mS|\tin}
		\nonumber\\&\qquad\qquad\qquad\qquad
				- \braket{\tout|\mS(N_\tin\cdot\ep^-)|\V p_1,\cdots,\V p_n}
			\Big]
		\\ &=
		\lim_{\w\to 0}(1+\w\p_\w)
			\Big[
				(N_\tout\cdot \ep^-)\braket{\V p_{m+1},\cdots,\V p_{m+n}|\mS|\tin}
		\nonumber\\&\qquad\qquad\qquad\qquad
				- (N_\tin\cdot\ep^-)\braket{\tout|\mS|\V p_1,\cdots,\V p_n}
			\Big],
			\label{equiv0}
\end{align}
where in the second equality we employ a convenient abuse of notation to write $N\cdot\ep^-$ both as an operator
	and as its action on the amplitude in the momentum-basis.
In section \ref{sec:equiv} we will see that, due to the presence of $\phi(\w_k)$,
	adding or removing subleading dressings do not change the value of the amplitude, that is,
\begin{align}
	\braket{\V p_{m+1},\cdots,\V p_{m+n}|\mS|\tin}=\braket{\tout|\mS|\V p_1,\cdots,\V p_n}=\braket{\tout|\mS|\tin}.
	\label{equiv}
\end{align}
This shows that the non-commutativity of subleading charges and the nonexistence of simultaneous eigenstates
	do not cause difficulties to S-matrix calculations at this order.
Using \eqref{equiv}, one may write \eqref{equiv0} as
\begin{align}
	\bra{\tout}[Q_S(\Yg),\mS]\ket{\tin}
		= \lim_{\w\to 0}(1+\w\p_\w)\left(N^\mn_\tout-N^\mn_\tin\right) \ep^-_\mn\braket{\tout|\mS|\tin}.
		\label{grav_Nout_minus_Nin}
\end{align}
Thus, the superrotation charge conservation \eqref{charge_conservation2} reads
\begin{align}
	\lim_{\w\to 0}(1+\w\p_\w)\left(N^\mn_\tout-N^\mn_\tin\right) \ep^-_\mn\braket{\tout|\mS|\tin} &=
		-iS_g^{(1)-}\braket{\tout|\mS|\tin},
\end{align}
which, with \eqref{subleading_softfactor}, implies for $\braket{\tout|\mS|\tin}\neq 0$,
\begin{align}
	\lim_{\w\to 0}(1+\w\p_\w)\left(N^\mn_\tout-N^\mn_\tin\right) \ep^-_\mn
		= -\sum_{i=1}^{m+n}\eta_i\frac{(p_i)^\mu k_\lambda (J_i)^{\lambda\nu}}{p_i\cdot k} \ep^-_\mn.
\end{align}
One can derive a similar relation associated with $\ep^+$ by choosing $Y=(\zb-\wb)^2(z-w)^{-1}\p_\zb$.
A natural split for the dressings is
\begin{align}
	\lim_{\w\to 0}(1+\w\p_\w)N^\mn_\tin &=
		-\sum_{i=1}^{m}\frac{(p_i)^\mu k_\lambda (J_i)^{\lambda\nu}}{p_i\cdot k},\\
	\lim_{\w\to 0}(1+\w\p_\w)N^\mn_\tout &=
		-\sum_{j=m+1}^{m+n}\frac{(p_j)^\mu k_\lambda (J_j)^{\lambda\nu}}{p_j\cdot k}.
\end{align}
If we treat supertranslation (associated with simple poles) separately as in \cite{Choi:2017ylo},
	we may assume that $N_{\tin,\tout}^\mn$ do not possess poles.
Then it follows that
\begin{align}
	N_\tin^\mn &= -\sum_{i=1}^{m}\frac{(p_i)^\mu k_\lambda (J_i)^{\lambda\nu}}{p_i\cdot k},\qquad
	N_\tout^\mn = -\sum_{j=m+1}^{m+n}\frac{(p_j)^\mu k_\lambda (J_j)^{\lambda\nu}}{p_j\cdot k},
\end{align}
which, substituted into \eqref{grav_linear}, yields the subleading soft Faddeev-Kulish dressings.
Put together with the leading soft gravitational Faddeev-Kulish dressings \cite{Ware:2013zja,Choi:2017bna,Choi:2017ylo},
	we denote the dressed asymptotic state with double brackets as
\begin{align}
	\kket{\V p_1,\cdots,\V p_n} = W_g(\V p_1,\cdots,\V p_n)\ket{\V p_1,\cdots,\V p_n},
	\label{dressed_state}
\end{align}
where $W_g(\V p_1,\cdots,\V p_n)$ is the gravitational $n$-particle dressing, which to the subleading order in soft
	momentum expansion and leading order in $\kappa$ is given by
\begin{align}
	W_g &= \exp\left\{
		\frac{\kappa}{2}
		\int\frac{d^3k}{(2\pi)^3}\frac{\phi(\w_k)}{2\w_k}
		\sum_{i=1}^n
		\frac{p_i^\mu p_i^\nu}{p_i\cdot k}
		(a_\mn^\dagger-a_\mn)
	\right\}
	\nonumber\\&\qquad\times
	\left(
		1-
		\frac{\kappa}{2}
		\int\frac{d^3k}{(2\pi)^3}\frac{\phi(\w_k)}{2\w_k}
		\sum_{i=1}^n
		\frac{p_i^\mu k_\rho J_i^{\rho\nu}}{p_i\cdot k}
		i(a_\mn^\dagger+a_\mn)
		+\mO(\kappa^2)
	\right),
	\label{dressing_raw}
\end{align}
which, keeping in mind that only order $\kappa$ terms can be trusted, may be conveniently written as\footnote{
	Non-commutativity of $p$ and $J$ may be ignored at this order.
	}
\begin{align}
	W_g &= \exp\left[
		\frac{\kappa}{2}
		\int\frac{d^3k}{(2\pi)^3}\frac{\phi(\w_k)}{2\w_k}
		\sum_{i=1}^n
		\frac{p_i^\mu}{p_i\cdot k}
		\left\{
		\left(p_i^\nu-ik_\rho J_i^{\rho\nu}\right)a_\mn^\dagger
		-\left(p_i^\nu+ik_\rho J_i^{\rho\nu}\right)a_\mn
		\right\}
		+\mO(\kappa^2)
	\right].
	\label{dressing0}
\end{align}
Here we employed the notation $a_\mn(\V k) = \sum_s \ep^{s*}_\mn(\V k)a_s(\V k)$, where $s$ spans all polarizations.
This includes unphysical polarizations, since the projection to physical polarizations in \eqref{subleading_softfactor}
	is a consequence of our choice \eqref{choice_grav} of $Y$; superrotation charge should be conserved for a generic vector field.
Unphysical polarizations are also required for canceling out infrared divergence, see \cite{Ware:2013zja,Choi:2017bna} for example.
The expression \eqref{dressing0} expanded to first order in $\kappa$ agrees with the gravitational Wilson line dressing of
	Mandelstam \cite{Mandelstam:1962us}.
Thus to this order in the coupling, one observes the equivalence between Wilson lines and FK dressings as discussed in \cite{Jakob:1990zi,Choi:2018oel}.

For explicit calculations, it is convenient to define the infrared function as
\begin{align}
	\phi(\w) = \begin{cases}
		\ 1\qquad \text{if $\cutoff < \w < \Esoft$},\\
		\ 0\qquad \text{otherwise},
	\end{cases}
	\label{irfunction}
\end{align}
where $\cutoff$ is the infrared cutoff which we take to be zero at the very end of the calculation, and $\Esoft$ is a very small energy scale
	below which particles are considered to be soft.
With this definition, the dressing \eqref{dressing0} becomes
\begin{align}
	W_g &= \exp\left[
		\frac{\kappa}{2}
		\int_{\cutoff<\w_k<\Esoft}\frac{d^3k}{(2\pi)^3}\frac{1}{2\w_k}
		\sum_{i=1}^n
		\frac{p_i^\mu}{p_i\cdot k}
		\left\{
		\left(p_i^\nu-ik_\rho J_i^{\rho\nu}\right)a_\mn^\dagger
		-\left(p_i^\nu+ik_\rho J_i^{\rho\nu}\right)a_\mn
		\right\}
		+\mO(\kappa^2)
	\right].
	\label{dressing}
\end{align}
The dressing $W_g$ acting on an $n$-particle Fock state will be understood as an $n$-particle dressing with the corresponding momenta of the hard particles,
	unless explicitly stated otherwise.
The dressed states automatically implement conservation of supertranslation charge \cite{Choi:2017bna} and superrotation charge, as shown above.

An important point concerning the validity of \eqref{dressing} should be emphasized here.
In any scattering process there are contributions from real emissions and from virtual diagrams.
The applicability of the subleading soft graviton and soft photon theorems to a $2\to 2$ scattering process has been studied
	in \cite{Luna:2016idw}.
There it was shown that the subleading soft photon theorem correctly reproduces the scattering amplitude to subleading order
	both for real and virtual photon emissions.
However, for the case of soft gravitons, the subleading soft graviton theorem correctly reproduces the real external emissions,
	but there is a violation of the theorem for virtual gravitons.
Thus, for the choice of vector field in \eqref{choice_grav}, we expect our dressing to be correct for the case of real emissions
	which we discuss in section \ref{sec:external}.
For scattering involving virtual gravitons we would need to generalize \eqref{dressing}.

\section{Dressing in QED} \label{sec:QED}

We present here the construction of subleading soft dressing in QED,
	which is fairly parallel to the case of gravity in section \ref{sec:gravity}.
In this section we will mostly follow the notation and conventions of \cite{Lysov:2014csa}.

\subsection{Mode expansions and conserved charges}

The photon field near $\mI^+$ becomes nearly free and can be approximated by
\begin{align}
	A_\mu^\tout(x) = e\sum_{\alpha=\pm}\int\frac{d^3k}{(2\pi)^3}\frac{1}{2\w_k}
		\left[
			\ep^{\alpha*}_\mu(\V k)a_\alpha^\tout(\V k) e^{ik\cdot x}
			+ \ep^{\alpha}_\mu(\V k)a_\alpha^{\tout\dagger}(\V k) e^{-ik\cdot x}
		\right],
	\label{mode_qedout}
\end{align}
where the out-operators satisfy the commutator
\begin{align}
	\left[a^\tout_\alpha(\V k),a^{\tout\dagger}_\beta(\V k')\right]=\delta_\ab (2\pi)^3(2\w_k)\delta^{(3)}(\V k-\V k').
\end{align}
Likewise, near $\mI^-$ we have the incoming photon field,
\begin{align}
	A_\mu^\tin(x) = e\sum_{\alpha=\pm}\int\frac{d^3k}{(2\pi)^3}\frac{1}{2\w_k}
		\left[
			\ep^{\alpha*}_\mu(\V k)a_\alpha^\tin(\V k) e^{ik\cdot x}
			+ \ep^{\alpha}_\mu(\V k)a_\alpha^{\tin\dagger}(\V k) e^{-ik\cdot x}
		\right],
	\label{mode_qedin}
\end{align}
with the standard commutator
\begin{align}
	\left[a^\tin_\alpha(\V k),a^{\tin\dagger}_\beta(\V k')\right]=\delta_\ab (2\pi)^3(2\w_k)\delta^{(3)}(\V k-\V k').
\end{align}

The QED analog of superrotation is the asymptotic symmetry on $\mI^\pm$ associated with Low's theorem.
In terms of the mode expansions \eqref{mode_qedout} and \eqref{mode_qedin}, the corresponding conserved charges are\footnote{
	We will use calligraphic font $\mQ$ for the QED charge to minimize notational overlap with gravity.} \cite{Lysov:2014csa}
\begin{align}
	\mQ^\pm_Y = \mQ_S^\pm(Y) + \mQ_H^\pm(Y),
	\label{charge_qed}
\end{align}
where the soft parts are given by 
\begin{align}
	\mQ^+_S(Y) &=
		-\frac{i}{4\pi e} \lim_{\w\to 0}(1+\w\p_\w)\int d^2z\,D_z^2Y^z \frac{\sqrt 2}{1+z\zb}
			\left[a^\tout_-(\w\V x_z)-a^{\tout\dagger}_+(\w\V x_z)\right]
		+ \hc,
	\label{soft_qed}
	\\
	\mQ^-_S(Y) &=
		\frac{i}{4\pi e} \lim_{\w\to 0}(1+\w\p_\w)\int d^2z\,D_z^2Y^z \frac{\sqrt 2}{1+z\zb}
		\left[a^\tin_-(\w\V x_z)-a^{\tin\dagger}_+(\w\V x_z)\right]
		+ \hc.
	\label{soft2_qed}
\end{align}
Here $\V x_z$ is a unit 3-vector whose direction is given by $(z,\zb)$; its Cartesian components are given in \eqref{x_z}.
The hard parts are defined by their actions on the Fock states,
\begin{align}
	\bra{\V p_1,\ldots,\V p_n} \mQ_H^+(Y) &=
		-i\sum_{i=1}^n Q_i
		\bigg(
			D_A Y^A(z_i)\p_{E_i}
			-\frac{1}{E_i}\mL_{Y(z_i)}
		\bigg)
		\bra{\V p_1,\ldots,\V p_n},
	\label{hard_qed}
	\\
	\mQ_H^-(Y) \ket{\V p_1,\ldots,\V p_n} &=
		i\sum_{i=1}^n Q_i
		\left(
			D_A Y^A(z_i)\p_{E_i}
			-\frac{1}{E_i}\mL_{Y(z_i)}
		\right)
		\ket{\V p_1,\ldots,\V p_n},
	\label{hard2_qed}
\end{align}
where $A\in\{z,\zb\}$, $Q_i$ is the electric charge of the $i$-th particle,
	and $\mL_Y$ is the Lie derivative on $S^2$, see appendix \ref{app:qed} for details.
Scattering processes conserve $\mQ_Y$, which implies that
\begin{align}
	\braket{\tout|(\mQ^+_Y\mS-\mS\mQ^-_Y)|\tin}=0,
	\label{charge_conservation_qed}
\end{align}
for any asymptotic states $\ket{\tin}$ and $\bra{\tout}$.

As was the case in gravity, the contribution to the subleading soft matrix element from a soft photon insertion
	$a_\alpha^{\tin\dagger}$ in the incoming state is equivalent to the contribution from an insertion $-a^\tout_{-\beta}$ in the outgoing state.
Therefore, we can follow the procedure of section \ref{sec:distinction} and define
\begin{align}
	\mQ_S(Y) \equiv -\frac{i}{4\pi e}\lim_{\w\to 0}(1+\w\p_\w)\int d^2z\,D_z^2Y^z\frac{\sqrt 2}{1+z\zb}
		\left[
			a_-(\w\V x_z)-a_+^\dagger(\w\V x_z)
		\right]+\hc,
	\label{soft3_qed}
\end{align}
where the soft operators are related by $\mQ_S^+ = \mQ_S = \left(\mQ_S^-\right)^\dagger = \mQ_S^-$, and
\begin{align}
	\left[a_\alpha(\V k),a_\beta^\dagger(\V k')\right]=\delta_\ab(2\pi)^3(2\w_k)\delta^{(3)}(\V k-\V k').
\end{align}
We emphasize that this procedure is done to avoid defining separate rules for contractions between operators on $\mI^+$ and $\mI^-$;
	one may obtain the same result with the distinction intact.

\subsection{Subleading soft dressing}

Now we construct the subleading soft Faddeev-Kulish dressing in QED to leading order in the coupling constant $e$
	as linearized coherent states that respect $\mQ_Y$ charge conservation \eqref{charge_conservation_qed}.

Using \eqref{charge_qed}-\eqref{hard2_qed}, \eqref{charge_conservation_qed} can be written as
\begin{align}
	\braket{\tout|\left[\mQ_S(Y),\mS\right]|\tin} &=
		i\sum_{i} Q_i
		\left(
			D_A Y^A(z_i)\p_{E_i}
			-\frac{1}{E_i}\mL_{Y(z_i)}
		\right)
		\braket{\tout|\mS|\tin}.
		\label{charge_conservation_qed2}
\end{align}
Let us choose
\begin{align}
	Y = \Ye \equiv \frac{(z-w)(1+z\zb)}{(\zb-\wb)}\p_z.
	\label{choice_qed}
\end{align}
Then, \eqref{charge_conservation_qed2} takes the form \cite{Lysov:2014csa} (see appendix \ref{app:qed} for details)
\begin{align}
	\braket{\tout|[\mQ_S(\Ye),\mS]|\tin} &=
		-\frac{\sqrt 2i}{e}S_e^{(1)-}\braket{\tout|\mS|\tin},
		\label{charge_conservation_qed3}
\end{align}
where $k^\mu \equiv (\w, \w\V x_z)$, and $S_e^{(1)-}$ is the subleading soft factor for negative-helicity photon,
\begin{align}
	S_e^{(1)-} = -ie\sum_i\eta_iQ_i\frac{k_\lambda J_i^{\lambda\nu}}{p_i\cdot k}\epsilon^-_\nu(\w\V x_z),
\end{align}
with $\eta_i=+1$ ($-1$) for outgoing (incoming) particle
	(by introducing $\eta_i$ we deviate from the convention of \cite{Lysov:2014csa}, see footnote \ref{eta_origin}).
An analogous expression involving $\ep^+$ can be derived by choosing $Y=(\zb-\wb)(1+z\zb)(z-w)^{-1}\p_\zb$ instead.
Using the identity $D_z^2\Ye^z = 2\pi(1+z\zb)\delta^{(2)}(z-w)$, one obtains the following expression for the soft charge,
\begin{align}
	\mQ_S(\Ye) &=
		-\frac{i}{\sqrt 2 e}\lim_{\w\to 0}(1+\w\p_\w)
		\left[a_-(\w\V x_z)-a_+^\dagger(\w\V x_z)\right].
\end{align}

Let us begin by considering a vacuum $\ket{0}$ such that,
\begin{align}
	\mQ_S(\Ye)\ket{0} \approx 0.
\end{align}
As noted in section \ref{sec:grav_subleading},
	formally the subleading soft charge does not annihilate the vacuum,
	but rather adds to it a soft photon.
As will be shown in section \ref{sec:external}, in scattering processes such a state completely factors out,
	and therefore in S-matrix computations one may act as if $Q_S$ annihilates the vacuum (hence the symbol $\approx$).

Now we consider states which are dressed to first order in $e$ that take the form
\begin{align}
	\ket{\tin} &=
		\left(
			1+ie\sum_{\alpha=\pm}\int\frac{d^3k}{(2\pi)^3}\frac{\phi(\w_k)}{2\w_k}
			\mN_\tin^\mu\left[
				\ep^{\alpha*}_\mu(\V k) a_\alpha(\V k)+\ep^\alpha_\mu(\V k) a_\alpha^\dagger(\V k)
			\right]
		\right)\ket{\V p_1,\cdots,\V p_m},
	\\
	\bra{\tout} &=
		\bra{\V p_{m+1},\cdots,\V p_{m+n}}
		\left(
			1-ie\sum_{\alpha=\pm}\int\frac{d^3k}{(2\pi)^3}\frac{\phi(\w_k)}{2\w_k}
			\mN_\tout^\mu\left[
				\ep^{\alpha*}_\mu(\V k) a_\alpha(\V k)+\ep^\alpha_\mu(\V k) a_\alpha^\dagger(\V k)
			\right]
		\right),
\end{align}
where $\mN_{\tin,\tout}^\mu$ are operators to be determined,
	and $\phi(\w_k)$ is the infrared function that restricts the momentum integrals to soft modes.
Then,
\begin{align}
	\mQ_S(\Ye)\ket{\tin} &=
		\frac{1}{\sqrt 2}\lim_{\w\to 0}(1+\w\p_\w)
			\sum_{\alpha=\pm}\int\frac{d^3k}{(2\pi)^3}\frac{\phi(\w_k)}{2\w_k}\mN_\tin^\mu
		\nonumber\\&\quad\times
		\left[
			a_-(\w\V x_z)-a_+^\dagger(\w\V x_z),
			\ep^{\alpha*}_\mu(\V k)a_\alpha(\V k)+\ep_\mu^\alpha(\V k)a_\alpha^\dagger(\V k)
		\right]\ket{\V p_1,\cdots,\V p_m}
	\\ &=		
		\frac{1}{\sqrt 2}\lim_{\w\to 0}(1+\w\p_\w)
			\sum_{\alpha=\pm}\int\frac{d^3k}{(2\pi)^3}\frac{\phi(\w_k)}{2\w_k}\mN_\tin^\mu
		\nonumber\\&\quad\times
		(2\pi)^3(2\w_k)\delta^{(3)}(\V k-\w\V x_z)
		\left(
			\ep^{\alpha*}_\mu(\V k)\delta_{\alpha,+}
			+ \ep^\alpha_\mu(\V k)\delta_{\alpha,-}
		\right)\ket{\V p_1,\cdots,\V p_m}
	\\ &=		
		\sqrt 2\lim_{\w\to 0}(1+\w\p_\w)
			\mN_\tin\cdot \ep^-(\w\V x_z)
			\ket{\V p_1,\cdots,\V p_m},
\end{align}
and
\begin{align}
	\bra{\tout}\mQ_S(\Ye) &=
		-\frac{1}{\sqrt 2}\lim_{\w\to 0}(1+\w\p_\w)
			\sum_{\alpha=\pm}\int\frac{d^3k}{(2\pi)^3}\frac{\phi(\w_k)}{2\w_k}
			\bra{\V p_{m+1},\cdots,\V p_{m+n}}\mN_\tout^\mu
		\nonumber\\&\quad\times
		\left[
			\ep^{\alpha*}_\mu(\V k)a_\alpha(\V k)+\ep_\mu^\alpha(\V k)a_\alpha^\dagger(\V k),
			a_-(\w\V x_z)-a_+^\dagger(\w\V x_z)
		\right]
	\\ &=		
		\sqrt 2\lim_{\w\to 0}(1+\w\p_\w)
			\bra{\V p_{m+1},\cdots,\V p_{m+n}}
			\mN_\tout\cdot \ep^-(\w\V x_z).
\end{align}
Along the same line of reasoning as in \eqref{grav_Nout_minus_Nin}, this leads to
\begin{align}
	\bra{\tout}[\mQ_S(\Ye),\mS]\ket{\tin}
		= \sqrt 2\lim_{\w\to 0}(1+\w\p_\w)\left(\mN^\mu_\tout-\mN^\mu_\tin\right) \ep^-_\mu\braket{\tout|\mS|\tin}.
\end{align}
Assume that the simple poles (associated with large gauge symmetry) have been treated separately, as in \cite{Kapec:2017tkm}.
Then $\lim_{\w\to 0}(1+\w\p_\w)\mN_{\tin,\tout}\cdot \ep^-=\mN_{\tin,\tout}\cdot \ep^-$,
	and the equation of charge conservation \eqref{charge_conservation_qed3} becomes
\begin{align}
	\left(\mN_\tout^\mu-\mN_\tin^\mu\right)\ep^-_\mu(\w\V x_z) &=
		-\sum_{i=1}^{m+n}\eta_iQ_i\frac{k_\rho J_i^{\rho\mu}}{p_i\cdot k}\epsilon^-_\mu(\w\V x_z),
\end{align}
if the matrix element $\braket{\tout|\mS|\tin}$ is to not vanish.
A natural splitting for the dressing is
\begin{align}
	\mN_\tin^\mu = -\sum_{i=1}^{m}Q_i\frac{k_\rho J_i^{\rho\mu}}{p_i\cdot k}.
	\qquad
	\mN_\tout^\mu = -\sum_{i=m+1}^{m+n}Q_i\frac{k_\rho J_i^{\rho\mu}}{p_i\cdot k},
\end{align}
Combining this with the known leading soft dressing, one deduces the dressed asymptotic state for QED to subleading order in the soft expansion
	and to first order in $e$ to be
\begin{align}
	\kket{\V p_1,\cdots,\V p_n} = W_e\ket{\V p_1,\cdots,\V p_n},
\end{align}
where the dressing $W_e$ is
\begin{align}
	W_e &= \exp\left\{
		e
		\int\frac{d^3k}{(2\pi)^3}\frac{\phi(\w_k)}{2\w_k}
		\sum_{i=1}^n
		\frac{Q_ip_i^\mu}{p_i\cdot k}
		(a_\mu^\dagger-a_\mu)
	\right\}
	\nonumber\\&\qquad\times
	\left(
		1-
		e
		\int\frac{d^3k}{(2\pi)^3}\frac{\phi(\w_k)}{2\w_k}
		\sum_{i=1}^nQ_i
		\frac{k_\rho J_i^{\rho\mu}}{p_i\cdot k}
		i(a_\mu^\dagger+a_\mu)
		+\mO(e^2)
	\right),
\end{align}
which, keeping in mind that only terms to first order in $e$ may be trusted, can be conveniently put as
\begin{align}
	W_e &=
		\exp\left[
			e\int\frac{d^3k}{(2\pi)^3}\frac{\phi(\w_k)}{2\w_k}
			\sum_{i=1}^n\frac{Q_i}{p_i\cdot k}
			\left\{
				\left(p_i^\mu - ik_\nu J_i^{\nu\mu}\right)a_\mu^\dagger
				-\left(p_i^\mu + ik_\nu J_i^{\nu\mu}\right)a_\mu
			\right\}
			+\mO(e^2)
		\right].
\end{align}
The term $\mO(e^2)$ emphasizes that the subleading dressing is valid only to order $e$.
The photon operator is defined as $a_\mu(\V k) = \sum_\alpha \ep^{\alpha*}_\mu(\V k)a_\alpha(\V k)$,
	where $\alpha$ spans all polarizations including unphysical ones, since the projection to physical polarizations is due to our choice \eqref{choice_qed} of $Y$;
	the charge should be conserved for a generic vector field.
The unphysical polarizations are also required to cancel out Weinberg's infrared-divergent factor.

With the explicit implementation \eqref{irfunction} of $\phi(\w_k)$, we obtain
\begin{align}
	W_e &=
		\exp\left[
			e\int_{\cutoff<\w_k<\Esoft}\frac{d^3k}{(2\pi)^3}\frac{1}{2\w_k}
			\sum_{i=1}^n\frac{Q_i}{p_i\cdot k}
			\left\{
				\left(p_i^\mu - ik_\nu J_i^{\nu\mu}\right)a_\mu^\dagger
				-\left(p_i^\mu + ik_\nu J_i^{\nu\mu}\right)a_\mu
			\right\}
			+\mO(e^2)
		\right],
\end{align}
where $\cutoff$ is the infrared cutoff and $\Esoft$ is the separation scale below which we consider particles to be soft.
Notice that the structure is very similar to the gravitational dressing \eqref{dressing}.
One can obtain the QED dressing from gravity by the replacement $\frac{\kappa}{2}(p_i^\mu \ep^s_\mn) \to eQ_i \ep^s_\nu$ for each particle.
The comments about non-commutativity of subleading and leading soft charges with all its complications discussed towards the
	end of section \eqref{sec:grav_subleading} also apply here.
In the subsequent sections, we will be working with the gravitational dressing with the understanding that
	same results can be shown for QED with minimal modifications.

\section{External soft gravitons and photons}\label{sec:external}

\begin{figure}[t]
	\centering
	\begin{subfigure}{.18\textwidth}
		\includegraphics[width=\textwidth]{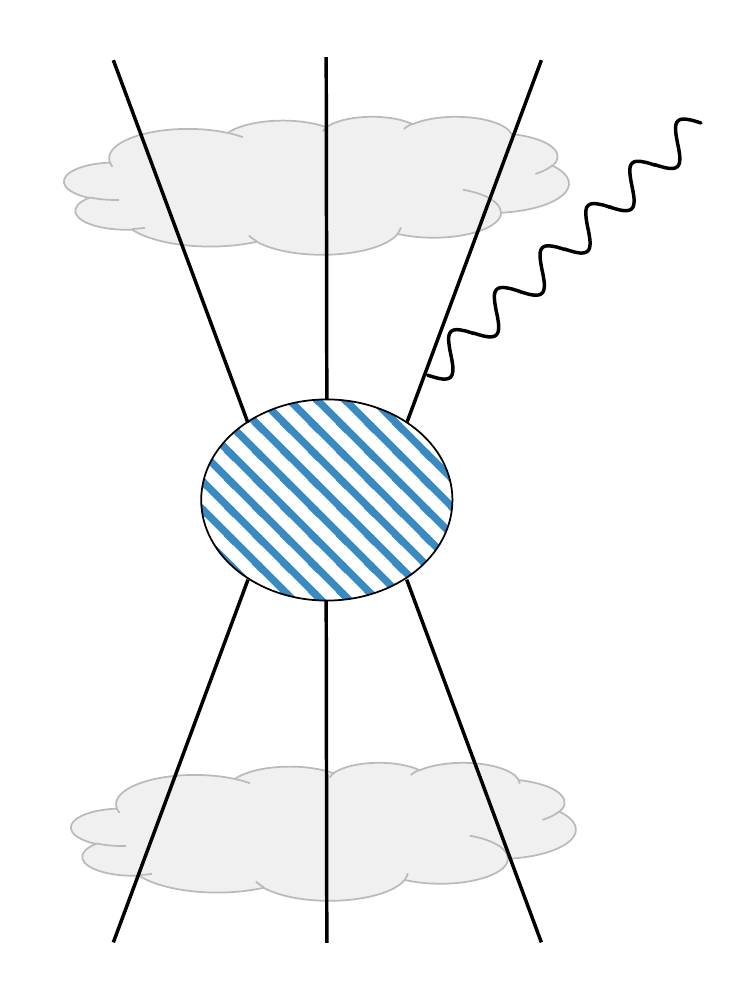}
		\caption{}
	\end{subfigure}
	\begin{subfigure}{.18\textwidth}
		\includegraphics[width=\textwidth]{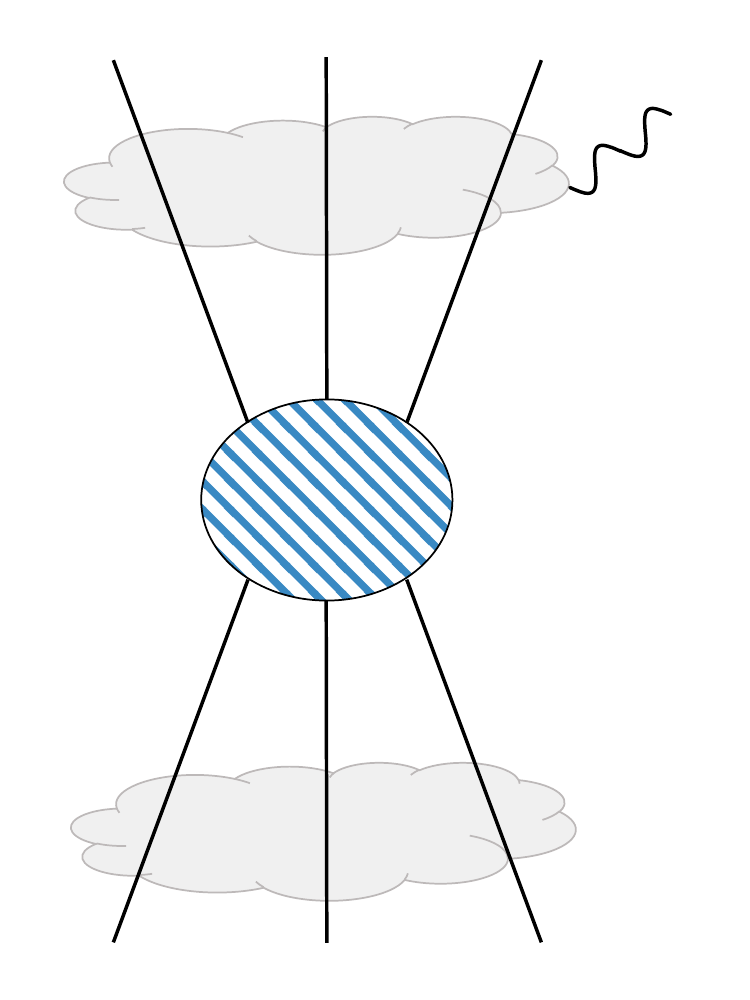}
		\caption{}
	\end{subfigure}
	\begin{subfigure}{.18\textwidth}
		\includegraphics[width=\textwidth]{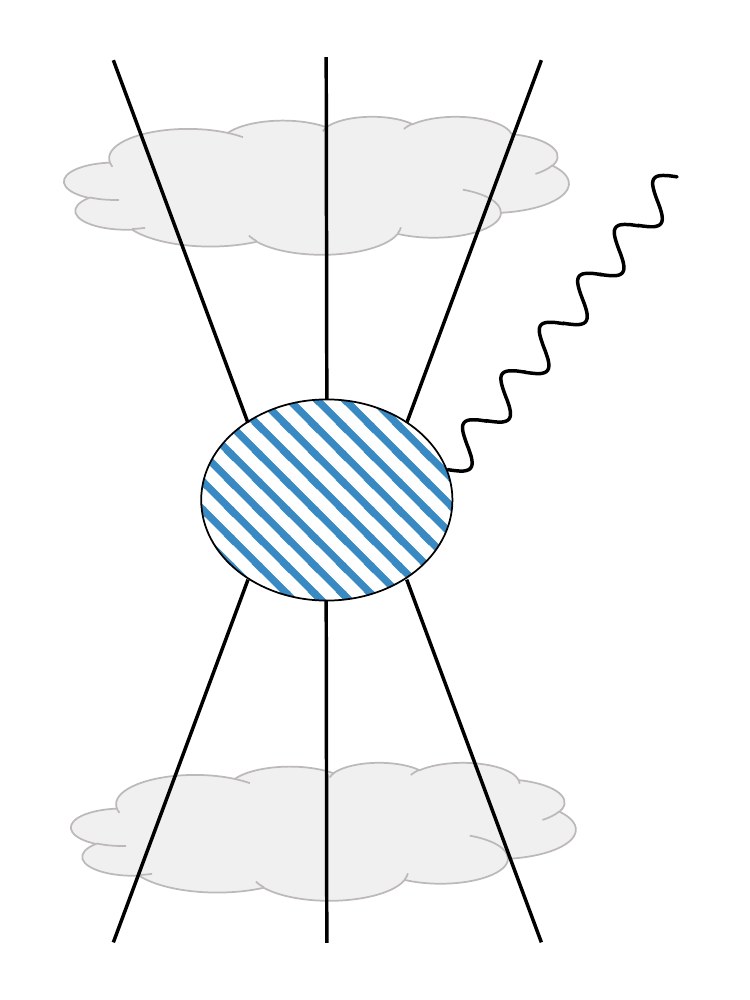}
		\caption{}
	\end{subfigure}
	\begin{subfigure}{.18\textwidth}
		\includegraphics[width=\textwidth]{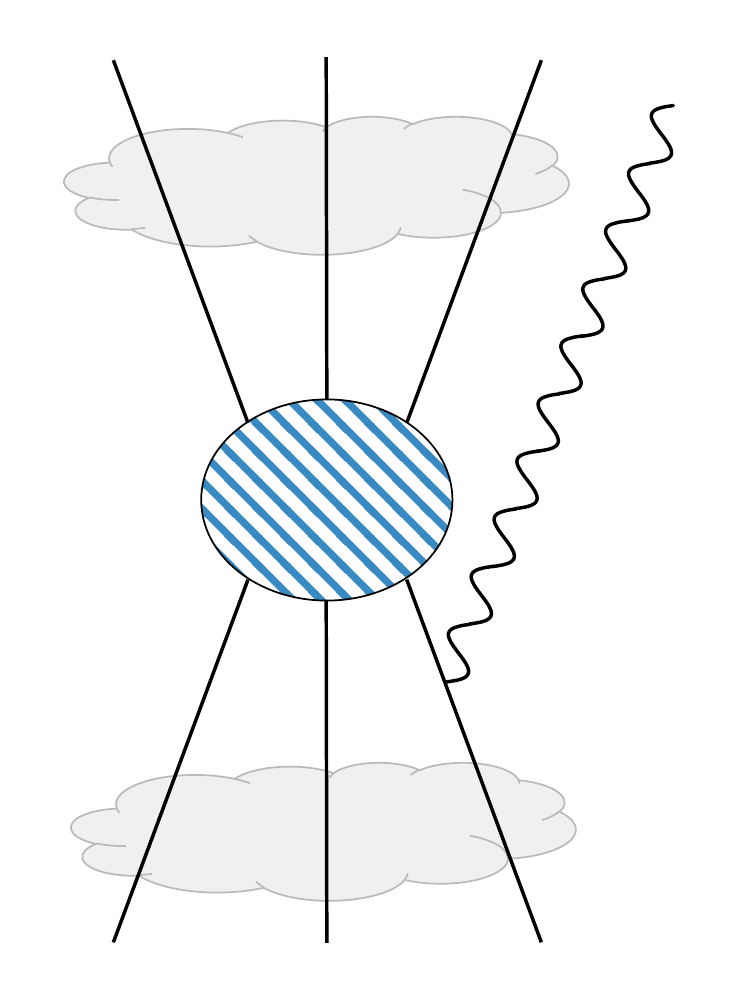}
		\caption{}
	\end{subfigure}
	\begin{subfigure}{.18\textwidth}
		\includegraphics[width=\textwidth]{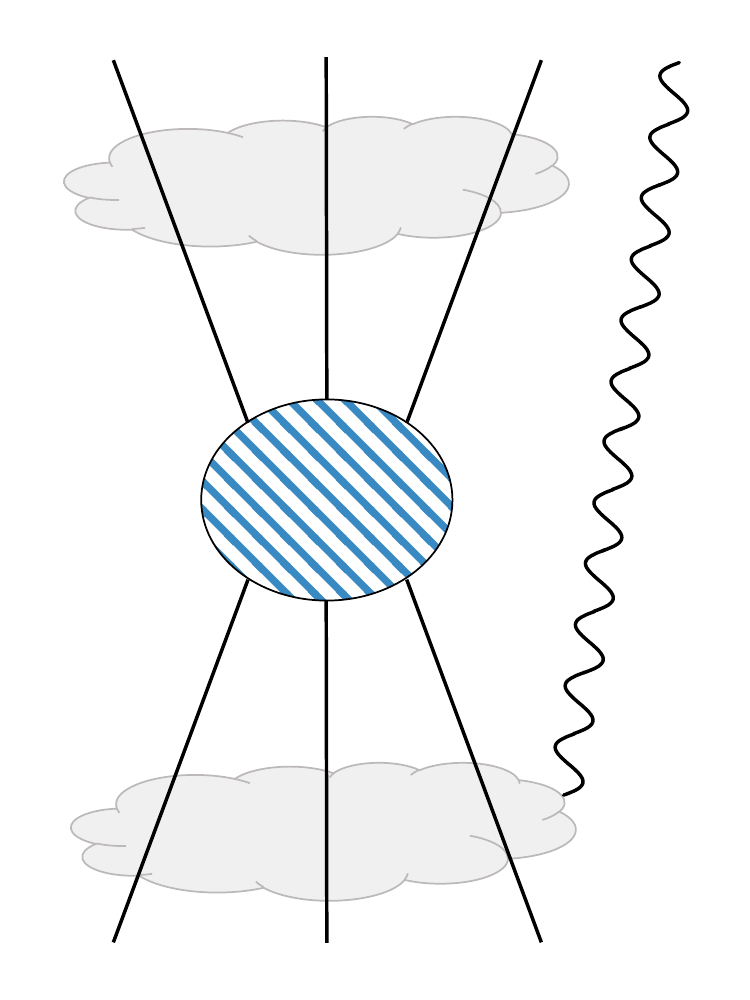}
		\caption{}
	\end{subfigure}
	\caption{Different contributions to the emission amplitude of a soft graviton.}
	\label{fig:emission}
\end{figure}

With the leading soft Faddeev-Kulish states, it is known that adding an external soft graviton does not induce infrared divergence;
	the divergent soft factors from the dressings cancel those from external legs \cite{Gabai:2016kuf, Choi:2017bna}.
Now that we have constructed dressings to subleading order in the soft expansion,
	we are in a position to investigate what happens to the $\mO(\w^0)$ subleading soft factors.
Although we will be working with gravitons, the derivation for QED is very similar and the final result is also valid for external soft photons.
Every step in the calculation can be changed to the corresponding expression for QED
	by replacing $\frac{\kappa}{2}(p_i^\mu\ep_\mn^s)$ with $eQ_i\ep_\nu^s$ for each hard particle.

As in \eqref{irfunction}, let $\Esoft$ be the soft energy scale, below which particles are considered to be soft,
	and let $\cutoff$ be the infrared cutoff which we take to be zero at the end of calculations.
Let us consider a scattering amplitude from the dressed $m$-particle state $\kket{\tin}$ to the dressed $n$-particle state $\kket{\tout}$,
	with a soft graviton insertion of polarization $s$ and momentum $k^\mu=(\w,\w\V x)$ where $\w$ is soft,
\begin{align}
	\mM\equiv \mM(k,s;\{p\})\equiv \bra{\tout} a_s(\w \V x) W_g^\dagger \mS W_g \ket{\tin},
	\qquad (\cutoff<\w<\Esoft).
\end{align}
The dressed amplitude $\mM$ has the small-$\w$ expansion
\begin{align}
	\mM(k,s;\{p\}) = \frac{1}{\w}\mM^{(-1)} + \mM^{(0)} + \mO(\w),
	\label{limM}
\end{align}
where each $\mM^{(n)}$ is independent of $\w$.
The different contributions to $\mM(k,s;\{p\})$ are illustrated in figure \ref{fig:emission}.

It is known that the first term involving the infrared-divergent amplitude $\mM^{(-1)}$ vanishes \cite{Choi:2017bna}.
To see this, note that $\mM^{(-1)}$ receives contribution from diagrams \ref{fig:emission}(a), \ref{fig:emission}(b),
	\ref{fig:emission}(d) and \ref{fig:emission}(e).
Using the notation $\eta_i=+1$ ($-1$) if $i$ is outgoing (incoming), we may write
\begin{align}
	\frac{1}{\w}\mM^{(-1)} &=
		\frac{\kappa}{2}
		\bigg[
			\underbrace{\sum_{i=1}^{m+n}\eta_i\frac{p_i^\mu p_i^\nu}{p_i\cdot k}}_\text{\ref{fig:emission}(a) and \ref{fig:emission}(d)}
			-\underbrace{\sum_{i=1}^{m+n}\eta_i\frac{p_i^\mu p_i^\nu}{p_i\cdot k}\phi(\w)}_\text{\ref{fig:emission}(b) and \ref{fig:emission}(e)}
		\bigg]\ep^s_\mn(\V k)\overline\mM = 0,
	\label{grav_leading}
\end{align}
where $\overline\mM\equiv \bbraket{\tout|\mS|\tin}=\braket{\tout|W_g^\dagger\mS W_g|\tin}$ is the dressed amplitude without graviton insertion.
In the second equation we used \eqref{irfunction} to write $\phi(\w)=1$.
The first sum in the square brackets comes from graviton emission from external legs (figures \ref{fig:emission}(a) and \ref{fig:emission}(d));
	the second sum comes from graviton emission from dressings (figures \ref{fig:emission}(b) and \ref{fig:emission}(e)).

We can determine $\mM^{(0)}$ by collecting the $\mO(\w^0)$ terms in the amplitude.
To do so, let us first decompose the gravitational dressing $W_g$ into leading and subleading parts,
\begin{align}
	W_g &= W^{(0)}_g W^{(1)}_g,\\
	W^{(0)}_g &=
		\exp\left\{
			\frac{\kappa}{2}
			\int\frac{d^3k}{(2\pi)^3}\frac{\phi(\w_k)}{2\w_k}
			\sum_{i}
			\frac{p_i^\mu p_i^\nu}{p_i\cdot k}
			(a_\mn^\dagger-a_\mn)
		\right\},\\
	W^{(1)}_g &=
		1-
		\frac{\kappa}{2}
		\int\frac{d^3k}{(2\pi)^3}\frac{\phi(\w_k)}{2\w_k}
		\sum_{i}
		\frac{p_i^\mu k_\rho J_i^{\rho\nu}}{p_i\cdot k}
		i(a_\mn^\dagger+a_\mn).
\end{align}
Then, we obtain the commutators
\begin{align}
	\left[a_s(\w\V x),W^\dagger_g(\V p_{m+1},\cdots,\V p_{m+n})\right]&=
		-\frac{\kappa}{2}\sum_{i=m+1}^{m+n}
		\left(
			\frac{p_i^\mu p_i^\nu}{p_i\cdot k}W^\dagger_g
			-i\frac{p_i^\mu k_\lambda J_i^{\lambda\nu}}{p_i\cdot k}W^{(0)\dagger}_g
		\right)\phi(\w)\epsilon_\mn^s,
	\label{com1}
	\\
	\left[a_s(\w\V x),W_g(\V p_1,\cdots,\V p_m)\right]&=
		\frac{\kappa}{2}\sum_{i=1}^{m}
		\left(
			W_g\frac{p_i^\mu p_i^\nu}{p_i\cdot k}
			-iW^{(0)}_g\frac{p_i^\mu k_\lambda J_i^{\lambda\nu}}{p_i\cdot k}
		\right)\phi(\w)\epsilon_\mn^s.
	\label{com2}
\end{align}
The first and second terms in the summands correspond respectively to the leading and subleading soft contributions
	from figures \ref{fig:emission}(b) and \ref{fig:emission}(e).	
The second terms contribute to $\mM^{(0)}$, along with emissions from internal propagators
	(figure \ref{fig:emission}(c)) and external legs (figures \ref{fig:emission}(a) and \ref{fig:emission}(d)).
There is one subtlety here -- the second terms in the summands are missing the subleading dressing factors $W^{(1)}_g$.
However, as we will see in section \ref{sec:equiv}, insertion of such factors only add $\mO(\Esoft)$ corrections to
	the amplitude, which is negligible by the definition of the soft energy scale $\Esoft$.
Therefore, within the amplitudes one may replace $W_g^{(0)}$ of \eqref{com1} and \eqref{com2} with $W_g$ and write
\begin{align}
	\mM^{(0)} &=
		\frac{\kappa}{2}
		\bigg[
			\underbrace{
				-i\sum_{i=1}^{m+n}\eta_i\frac{p^\mu_ik_\lambda J_i^{\lambda\nu}}{p_i\cdot k}
			}_\text{\ref{fig:emission}(a), \ref{fig:emission}(c) and \ref{fig:emission}(d)}
			+ \underbrace{
				i\sum_{i=1}^{m+n}\eta_i\frac{p^\mu_ik_\lambda J_i^{\lambda\nu}}{p_i\cdot k}\phi(\w)
			}_\text{\ref{fig:emission}(b) and \ref{fig:emission}(e)}
		\bigg]\ep^s_\mn(\V k)\overline\mM
		=\mO(\Esoft),
	\label{grav_subleading}
\end{align}
since $\phi(\w)=1$ for $\w<\Esoft$.
We remind the reader that the sign $\eta_i$ in the subleading soft factor comes from
	the different momentum-space representations of the action of $J_i^\mn$ on bras and kets:
\begin{align}
	\bra{\V p}J^\mn &= -i\left(p^\mu\frac{\p}{\p p_\nu} - p^\nu \frac{\p}{\p p_\mu}\right)\bra{\V p},\\
	J^\mn\ket{\V p} &= i\left(p^\mu\frac{\p}{\p p_\nu} - p^\nu \frac{\p}{\p p_\mu}\right)\ket{\V p},
\end{align}
which may differ from some conventions in the literature, see footnote \ref{eta_origin}.

Collecting the results \eqref{grav_leading} and \eqref{grav_subleading}, equation \eqref{limM} becomes
\begin{align}
	\mM(k,s;\{p\})=\underbrace{\frac{1}{\w}\mM^{(-1)}}_{=0} + \underbrace{\vphantom{\frac{1}{\w}}\mM^{(0)}}_{=\mO(\Lambda)} + \mO(\w)=\mO(\Lambda).
	\label{negligible}
\end{align}
At this point one may remove the infrared regulator $\cutoff\to 0$, and conclude that the soft emission amplitude
	is negligible since $\w$ is by definition less than the soft energy scale $\Esoft$, which in turn is by definition much less
	than any energy scales of our interest.
As the emission amplitude vanishes in the soft limit,
	the state containing a zero-energy graviton can be treated as null as far as scattering processes are concerned:
\begin{align}
	\lim_{\w\to 0}a^\dagger_s(\w\V x)\ket{0}\approx 0.
\end{align}

In summary, the use of leading and subleading Faddeev-Kulish states
	do not allow absorption and emission of on-shell soft gravitons at tree level.

\section{Equivalence of Faddeev-Kulish amplitudes and traditional amplitudes} \label{sec:equiv}

In this section, we show that the Faddeev-Kulish amplitude is equivalent to the infrared-finite part of traditional amplitudes
	constructed using Fock states, up to power-law type corrections in the soft energy scale $\Esoft$ which is negligible by definition.
Keeping both leading and subleading terms to first order in $\kappa$ in the exponent of the dressing function $W_g$, we explicitly show that this equivalence is
	up to order $\Lambda$ for radiation-less amplitudes in the case of scattering of a scalar from an external potential.
In reference \cite{Choi:2017bna} only the infrared-finiteness was shown, keeping just the leading order term in the exponent of the dressing.
Again, although we will only derive the result explicitly for gravity, the derivation for QED is similar and the result also holds for QED amplitudes.

For simplicity we will consider a $1\to 1$ gravitational potential scattering
	between the dressed states $\kket{\V p_i}$ and $\bbra{\V p_f}$ at one-loop order.
Let us define the shorthand notation
\begin{align}	
	P_i^\mn\equiv \frac{\kappa}{2}\left(\frac{p^\mu_i p^\nu_i}{p_i\cdot k}\right),
	\quad
	Q_i^\mn\equiv \frac{\kappa}{2}\left(-i\frac{p^\mu_i k_\rho J_i^{\rho \nu}}{p_i\cdot k}\right),
\end{align}
and similarly $P_f^\mn$ and $Q_f^\mn$ corresponding to $p_f$.
We also use the notation
\begin{align}
	\int\td{k} \equiv \int_{\cutoff<\w_k<\Esoft} \frac{d^3k}{(2\pi)^3}\frac{1}{2\w_k},
\end{align}
where $\cutoff$ is the infrared cutoff and $\Esoft$ is the soft energy scale.
We can use these and \eqref{dressed_state} to write the dressed states as
\begin{align}
	\bbra{\V p_f}
		&= \bra{\V p_f}W^\dagger_g(\V p_f)
		= \bra{\V p_f}
		\left(
			1-\int\td{k}Q_f a
		\right)
		\exp\left\{
			-\int\td{k}	P_f(a^\dagger-a)
		\right\}
	, \\
	\kket{\V p_i}
		&= W_g(\V p_i)\ket{\V p_i}
		=
		\exp\left\{
			\int\td{k}P_i(a^\dagger-a)
		\right\}
		\left(
			1+\int\td{k}Q_i a^\dagger
		\right)
		\ket{\V p_i}
	,
\end{align}
where concatenation implies contraction, for example,
\begin{align}
	P_i(a^\dagger-a) \equiv P^\mn_i(a_\mn^\dagger-a_\mn).
\end{align}

\begin{figure}[t]
	\centering
	\begin{subfigure}{.16\textwidth}
		\includegraphics[width=\textwidth]{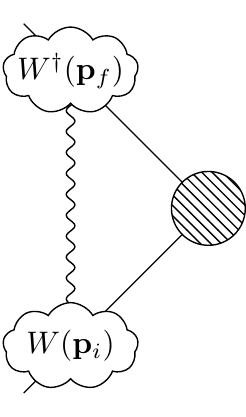}
		\caption{}
	\end{subfigure}
	\qquad
	\begin{subfigure}{.16\textwidth}
		\includegraphics[width=\textwidth]{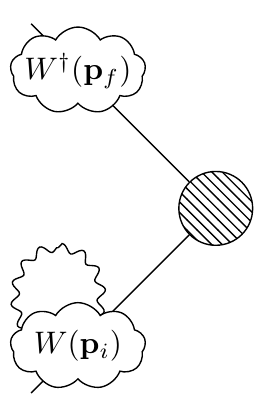}
		\caption{}
	\end{subfigure}
	\qquad
	\begin{subfigure}{.16\textwidth}
		\includegraphics[width=\textwidth]{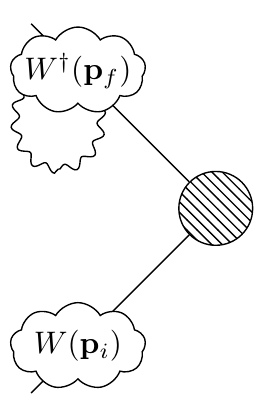}
		\caption{}
	\end{subfigure}
	\caption{
		We consider the simple case of $1\to 1$ gravitational potential scattering,
			where the incoming and outgoing momenta are $\V p_i$ and $\V p_f$, respectively.
		The figures illustrate different contributions to the FK amplitude of this process.
		Blob represents the internal diagram, including the gravitational potential.
		}
	\label{fig:cloud}
\end{figure}

First consider the contribution to the matrix element due to graviton exchange between dressings.
There are two self-interactions of the dressings, each coming from $W_g(\V p_i)$ and $W_g^\dagger(\V p_f)$
	(figures \ref{fig:cloud}(b) and \ref{fig:cloud}(c)),
	and one cross-interaction between $W_g(\V p_i)$ and $W_g^\dagger(\V p_f)$ (figure \ref{fig:cloud}(a)).
Using the Baker-Campbell-Hausdorff (BCH) formula,
\begin{align}
	e^{A+B}=e^Ae^Be^{-\frac{1}{2}[A,B]}\qquad\text{if $[A,[A,B]]=[B,[A,B]]=0$},
\end{align}
we may write,
\begin{align}
	\exp\left\{
		-\int\td{k} P_f(a^\dagger-a)
	\right\}
	=
	\exp\left\{
		-\int\td{k} P_f a^\dagger
	\right\}	
	\exp\left\{
		\int\td{k} P_f a
	\right\}	
	\exp\left\{
		-\frac{1}{4}\int\td{k} P_f I P_f
	\right\}	
\end{align}
and similar for the incoming dressing, where we used (see \cite{Ware:2013zja,Donoghue:2017pgk} for example)
\begin{align}
	\left[a_\mn(\V k),a_\rs^\dagger(\V k')\right] &= \frac{1}{2}I_\mnrs (2\pi)^3(2\w_k)\delta^{(3)}(\V k-\V k'),
	\\
	I_\mnrs &= \eta_{\mu\rho}\eta_{\nu\sigma}+\eta_{\mu\sigma}\eta_{\nu\rho}-\eta_\mn\eta_\rs,
\end{align}
and employed the notation $P_f I P_f\equiv P_f^\mn I_\mnrs P^\rs_f$.
Then to first order in $\kappa$,\footnote{
While we write the dressings to first order in $\kappa$, we will keep the $\kappa^2$-order infrared-divergent term $\int\td{k} P_f I P_f$
	since it is needed to cancel infrared divergence \cite{Ware:2013zja}.
}
\begin{align}
	\bbra{\V p_f}
		&= \bra{\V p_f}
		\left(
			1-\int\td{k}Q_f a+\int\td{k} P_f a-\frac{1}{4}\int\td{k} P_f I P_f
		\right)
	\label{self1}
	, \\
	\kket{\V p_i}
		&=
		\left(
			\int\td{k}P_i a^\dagger
			+\int\td{k}Q_i a^\dagger
			-\frac{1}{4}\int\td{k}P_iIP_i
		\right)
		\ket{\V p_i}
	\label{self2}
	.
\end{align}
The last terms on the RHS of \eqref{self1} and \eqref{self2} are the self-interaction contributions of the dressings to the matrix element,
\begin{align}
	-\frac{1}{4}\int\td{k} (P_fIP_f+P_iIP_i)\braket{\V p_f|\mS|\V p_i}.
	\label{cloud1}
\end{align}
The cross-interaction between the two dressings (figure \ref{fig:cloud}(a))
	come from the contraction between the graviton operators of \eqref{self1} and \eqref{self2}, which introduces the term
\begin{align}
		\frac{1}{2}\int\td{k}\left(P_fIP_i -Q_fIP_i +Q_iIP_f\right)
		\braket{\V p_f|\mS|\V p_i}.
	\label{cloud2}
\end{align}

\begin{figure}[t]
	\centering
	\begin{subfigure}{.16\textwidth}
		\includegraphics[width=\textwidth]{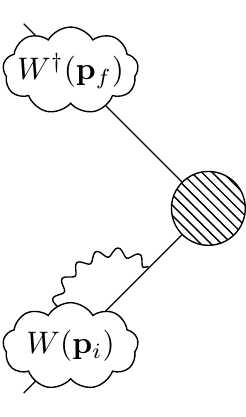}
		\caption{}
	\end{subfigure}
	\begin{subfigure}{.16\textwidth}
		\includegraphics[width=\textwidth]{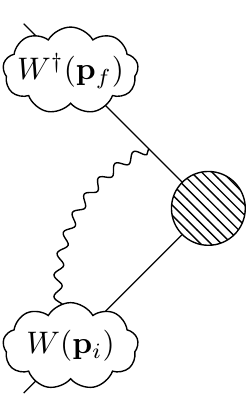}
		\caption{}
	\end{subfigure}
	\begin{subfigure}{.16\textwidth}
		\includegraphics[width=\textwidth]{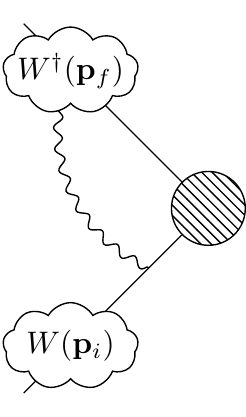}
		\caption{}
	\end{subfigure}
	\begin{subfigure}{.16\textwidth}
		\includegraphics[width=\textwidth]{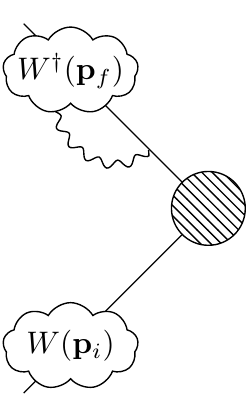}
		\caption{}
	\end{subfigure}
	\begin{subfigure}{.16\textwidth}
		\includegraphics[width=\textwidth]{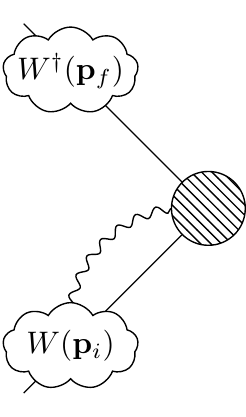}
		\caption{}
	\end{subfigure}
	\begin{subfigure}{.16\textwidth}
		\includegraphics[width=\textwidth]{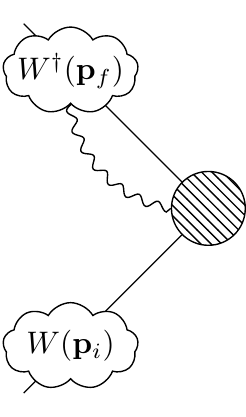}
		\caption{}
	\end{subfigure}
	\caption{
		Contributions to the amplitude due to graviton exchange between a dressing and either an external leg or the internal diagram.
		}
	\label{fig:amplitude}
\end{figure}

Now, let us consider the remaining contributions, namely the ones due to a dressing exchanging gravitons with
	either an external leg or an internal propagator, see figure \ref{fig:amplitude}.
At the leading soft (divergent) order, graviton exchange with an internal propagator does not contribute, while each
	exchange with an external leg (figures \ref{fig:amplitude}(a)-(d))
	induces a soft factor $\pm\eta P^\mn$, where the $+$ ($-$) sign corresponds
	to emission (absorption) of the graviton, and $\eta=1$ ($-1$) if the external leg is outgoing (incoming).
At the next order, graviton exchanges with the internal diagram (figures \ref{fig:amplitude}(e) and \ref{fig:amplitude}(f))
	induce one subleading soft factor $\eta Q^\mn + \tilde Q^\mn$ for each external leg, where $\tilde Q^\mn=\mO(\w^0)$ is a subleading soft
	factor due to the graviton being off-shell; see \cite{Luna:2016idw} for example.
Since there are two dressings and two external legs, at one-loop level this introduces eight terms,
	four from $W^\dagger(\V p_f)$ which correspond to diagrams \ref{fig:amplitude}(c), \ref{fig:amplitude}(d) and \ref{fig:amplitude}(f),
\begin{align}
		\frac{1}{2}\int\td{k}
		\Big[
			(P_f-Q_f)IP_f
			+(Q_f+\tilde Q_f)IP_f
			-(P_f-Q_f)IP_i
			+(-Q_i+\tilde Q_i)IP_f
		\Big]
	\braket{\V p_f|\mS|\V p_i},
	\label{cloud3}
\end{align}
and four from $W(\V p_i)$ corresponding to figures \ref{fig:amplitude}(a), \ref{fig:amplitude}(b) and \ref{fig:amplitude}(e),
\begin{align}
		\frac{1}{2}\int\td{k}
		\Big[
			-(P_i+Q_i)IP_f
			+(Q_f+\tilde Q_f)IP_i
			+(P_i+Q_i)IP_i
			+(-Q_i+\tilde Q_i)IP_i
		\Big]
	\braket{\V p_f|\mS|\V p_i}.
	\label{cloud4}
\end{align}

Expressions \eqref{cloud1}, \eqref{cloud2}, \eqref{cloud3} and \eqref{cloud4} comprise the full contribution of the graviton interaction
	to the matrix element that involves Faddeev-Kulish dressings at one-loop.
The net contribution to subleading order in soft momentum is
\begin{align}
	\frac{1}{4}\int\td{k}\Big[
			(P_f-P_i)I(P_f-P_i)
			-2(Q_i-\tilde Q_i-\tilde Q_f)IP_f
			+2(Q_f+\tilde Q_i+\tilde Q_f)IP_i
		\Big]\braket{\V p_f|\mS|\V p_i}.
	\label{term1}
\end{align}

\begin{figure}[t]
	\centering
	\begin{subfigure}{.16\textwidth}
		\includegraphics[width=\textwidth]{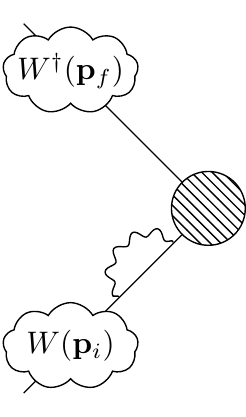}
		\caption{}
	\end{subfigure}
	\begin{subfigure}{.16\textwidth}
		\includegraphics[width=\textwidth]{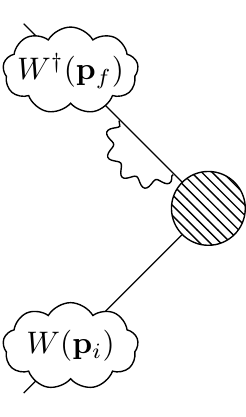}
		\caption{}
	\end{subfigure}
	\begin{subfigure}{.16\textwidth}
		\includegraphics[width=\textwidth]{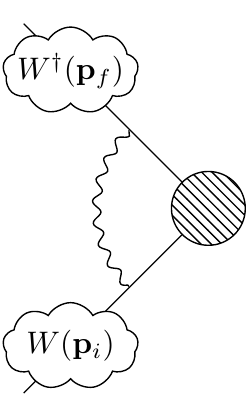}
		\caption{}
	\end{subfigure}
	\begin{subfigure}{.16\textwidth}
		\includegraphics[width=\textwidth]{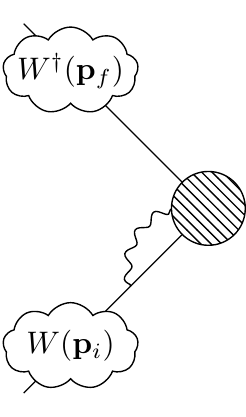}
		\caption{}
	\end{subfigure}
	\begin{subfigure}{.16\textwidth}
		\includegraphics[width=\textwidth]{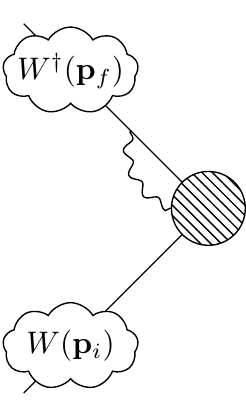}
		\caption{}
	\end{subfigure}
	\caption{
		Graviton exchanges that do not involve dressings.
		The soft graviton loops exponentiate and factor out \cite{Weinberg:1965nx}.
		}
	\label{fig:divergent}
\end{figure}

Now, the traditional amplitude $\braket{\V p_f|\mS|\V p_i}$ is infrared divergent, owing to the presence of soft virtual graviton loops.
The divergence can be factored out, such that at one loop \cite{Weinberg:1965nx,Choi:2017bna},
\begin{align}
	\braket{\V p_f|\mS|\V p_i} &=
		\left(1-\frac{1}{4}\int\td{k}(P_f-P_i)I(P_f-P_i)\right)
		\overline{\braket{\V p_f|\mS|\V p_i}},
	\label{term2}
\end{align}
where $\overline{\braket{\V p_f|\mS|\V p_i}}$ is the infrared-finite matrix element
	where all virtual graviton loop momenta below the soft energy scale $\Esoft$ has been removed.
In figure \ref{fig:divergent}, only diagrams \ref{fig:divergent}(a)-(c) contribute to this term.
We will not concern ourselves with subleading corrections to this factor coming from diagrams \ref{fig:divergent}(a)-(e),
	since such corrections do not alter our conclusion.

Putting \eqref{term1} and \eqref{term2} together, at one-loop we observe that
\begin{align}
	\bbraket{\V p_f\Vert\mS\Vert\V p_i} &=
		\left(1+
			\frac{1}{2}\int\td{k}\left[
				(Q_f+\tilde Q_i+\tilde Q_f)IP_i
				-(Q_i-\tilde Q_i-\tilde Q_f)IP_f
			\right]
		\right)
		\overline{\braket{\V p_f|\mS|\V p_i}}.
		\label{semifinal}
\end{align}
Recall that the integral spans only the soft sector $\cutoff<\w_k<\Esoft$.
Since the integrand is of order $\mO(\w_k^{0})$, after removing the
	infrared regulator $\cutoff$ the second term on the RHS of \eqref{semifinal} becomes
\begin{align}
	\frac{1}{2}\int\limits_{0<\w_k<\Esoft}\frac{d^3k}{(2\pi)^3}\frac{1}{2\w_k}\left[
		(Q_f+\tilde Q_i+\tilde Q_f)IP_i
		-(Q_i-\tilde Q_i-\tilde Q_f)IP_f
	\right]\overline{\braket{\V p_f|\mS|\V p_i}} &= \mO(\Esoft),
	\label{Lambda}
\end{align}
By definition, $\Esoft$ is the soft scale and is therefore negligible compared to all other energy scales of significance.
Therefore,
\begin{align}
	\bbraket{\V p_f\Vert\mS\Vert\V p_i} &=
		\overline{
		\braket{\V p_f|\mS|\V p_i}
		}.
	\label{dress_trad}
\end{align}
That is, the Faddeev-Kulish amplitude dressed to subleading soft order is equivalent to the infrared-finite traditional matrix element.
This extends the results of \cite{Ware:2013zja,Choi:2017bna} where the analyses were done only at the level of infrared-divergent terms.

Now, recall that our construction of dressings does not account for loop-corrections to the subleading soft theorems.
As one can see from \eqref{charge_conservation2} and \eqref{charge_conservation_qed3},
	such corrections will introduce $\ln \w$ terms to the integrand in the exponent of the dressings.
Then, from \eqref{Lambda} we observe that the corresponding corrections to dressed amplitudes will
	involve infrared-finite integrals of the form
\begin{align}
	\lim_{\cutoff\to 0}\int_\cutoff^{\Esoft}d\w_k\,\ln\w_k &= \mO(\Esoft \ln \Esoft),
\end{align}
which becomes arbitrarily small as we decrease $\Esoft$, and therefore is negligible compared to other
	energy scales of our interest.
It follows that the dressed amplitudes remain unaffected by loop corrections to the subleading soft theorem.

One should check that the equivalence \eqref{dress_trad} reproduces the inclusive cross section obtained by
	the Bloch-Nordsieck method \cite{BN}, which sums over emissions of real gravitons with energies below the detector resolution $\Eres$.
It was shown in \cite{Weinberg:1965nx} that the cross section for a transition $\alpha\to\beta$ accompanied by
	any number of real gravitons with total energy below $\Eres$ can be written as
\begin{align}
	\Gamma_{\alpha\to\beta}(\leq \Eres) &=
		\frac{1}{\pi}\sum_{N=0}^\infty \int_{\cutoff}^\Eres d\w_1\cdots\int_\cutoff^\Eres d\w_N
		\int_{-\infty}^\infty d\sigma\frac{\sin (\Eres\sigma)}{\sigma}
		\nonumber\\&\quad\times\exp\left \{i\sigma\sum_{i=1}^N\w_i\right \}
		\Gamma_{\alpha\to\beta}(\w_1,\cdots,\w_N),
	\label{cross}
\end{align}
where $\Gamma_{\alpha\to\beta}(\w_1,\cdots,\w_N)$ denotes the cross section for emission of $N$ gravitons with
	energies $\w_1,\cdots,\w_N$.
The cross sections are now the norm-squared of the dressed amplitudes.
In the previous section, we have shown in \eqref{negligible} that scattering amplitudes, and therefore cross sections, for processes
	that emit/absorb real gravitons with energy below the soft scale $\Esoft$ are negligible.
This has the effect of replacing the infrared cutoff $\cutoff$ in \eqref{cross} with the soft energy scale $\Esoft$,
	which results in \cite{Weinberg:1965nx}
\begin{align}
	\Gamma_{\alpha\to\beta}(\leq \Eres) &= \left(\frac{\Eres}{\Esoft}\right)^Bb(B)\Gamma_{\alpha\to \beta},
\end{align}
where $b(x)= 1-\frac{1}{12}\pi^2x^2+\cdots$ and
\begin{align}
	B &=
		\frac{\kappa^2}{64\pi^2}\sum_{ij}\eta_i\eta_j \frac{m_im_j(1+\beta_{ij}^2)}{\beta_{ij}(1-\beta_{ij}^2)^{1/2}}
			\ln\left(\frac{1+\beta_{ij}}{1-\beta_{ij}}\right),
		\qquad
		\beta_{ij}^2\equiv 1-\frac{m_i^2m_j^2}{(p_i\cdot p_j)^2}.
\end{align}
$\Gamma_{\alpha\to\beta}$ is the cross section for $\alpha\to \beta$ without the undetectable real gravitons.
In the original construction of \cite{Weinberg:1965nx} with Fock states, one factors out the soft loop contribution from $\Gamma_{\alpha\to\beta}$.
In our construction with the dressed amplitude \eqref{dress_trad}, we have no soft graviton loops
	(they have been canceled on account of the dressings) and thus may write
\begin{align}
	\Gamma_{\alpha\to\beta}=\Gamma_{\alpha\to\beta}^0,
\end{align}
where $\Gamma_{\alpha\to\beta}^0$ is the cross section computed excluding virtual graviton loop momenta below the soft scale $\Esoft$.
Therefore,
\begin{align}
	\Gamma_{\alpha\to\beta}(\leq \Eres) &= \left(\frac{\Eres}{\Esoft}\right)^Bb(B)\Gamma_{\alpha\to \beta}^0,
\end{align}
which agrees with the inclusive cross section computed in \cite{Weinberg:1965nx} using Fock states via the Bloch-Nordsieck method.

{\color{red}
}

\section{Discussions}

Using ideas similar to the one presented in \cite{Choi:2017ylo}, at leading order in the coupling
	we constructed the Faddeev-Kulish dressing for gravity and QED to subleading order in the soft energy expansion.
We have shown that the dressed amplitudes are equivalent to the infrared-finite part
	of the traditional matrix elements, up to negligible power-law type corrections of the soft energy scale.
We have also shown that, to first-order in the coupling constant,
	the FK state formalism does not allow soft radiation of photon and graviton.
This supports the proposition that for the FK states soft particles carry information about the hard particles and vice versa
	\cite{Carney:2017jut,Carney:2017oxp}.

An important question that we have not addressed in this paper is the characterization of a general dressed state up to and including
	all subleading soft corrections.
Since the subleading charges do not commute with each other or with the leading charge, it is not clear how to construct dressed states
	beyond the order $\kappa$ term in the exponent of the dressing operator.
Another manifestation of this in perturbation theory is that the subleading corrections do not exponentiate \cite{Luna:2016idw,Akhoury:2013yua}.
This connection between the non-Abelian nature of the asymptotic symmetry charges and the violation of factorization of soft quanta has not been
	noticed before.
Indeed, in non-Abelian gauge theories there is a breakdown of exponentiation of soft gluon effects for certain processes even at the leading soft order.
Further studies of their connection may enhance our understanding of both of these problems.
See reference \cite{Gonzo:2019fai} for a recent work on the leading soft FK dressings and asymptotic symmetries of perturbative QCD.

We end with some comments about the loop corrections.
The subleading theorem is known to receive $\ln \w$ corrections at one-loop level,
	see \cite{Bern:2014oka,He:2014bga, AtulBhatkar:2018kfi, Sahoo:2018lxl, Campiglia:2019wxe} for example.
Since we have not considered such corrections in our construction, this is another reason why the subleading soft dressings we derived
	may be trusted only to first order in the coupling constants.
We have argued that the agreement of FK amplitudes with traditional amplitudes is not affected by the correction.
However, the non-existence of soft radiation presented in section \ref{sec:external} is a tree-level statement.
It would be interesting to study how loop corrections affect this result.

\acknowledgments

We are very grateful to Andy Strominger for insightful comments on the first draft of the manuscript.
We would also like to thank Temple He for helpful comments.
SC thanks the organizers and participants of the Quantum Gravity and Quantum Information workshop
	at CERN in March 2019 for discussions that in part motivated this work.
He also acknowledges Rackham Graduate School for the travel support.
SC is supported by the Samsung Scholarship and the Leinweber Graduate Fellowship.

\appendix

\section{Subleading soft factors and spin angular momenta}\label{app:ssfnsam}

In this section, we review the steps presented in \cite{Kapec:2014opa,Lysov:2014csa}
	of deriving the subleading soft factors from the action of hard charge on matter particles.
We will treat gravity with massless scalars first, and then examine how the presence of spin affects the result.
This will then be used to derive analogous results for QED.

\subsection{Gravity}\label{app:grav}

\subsubsection{Massless scalars}

From the actions of soft and hard superrotation charges \cite{Kapec:2014opa}, we have
\begin{align}
	\braket{\tout|[Q_H,\mS]|\tin} &=
		i\sum_i \left(
			Y^z(z_i)\p_{z_i}-\frac{E_i}{2}D_zY^z(z_i)\p_{E_i}+\zz
		\right)\braket{\tout|\mS|\tin},
	\label{hard1}
	\\
	\braket{\tout|[Q_S,\mS]|\tin} &=
		-\frac{i}{2\pi\kappa}\lim_{\w\to 0}(1+\w\p_\w)\int d^2zD_z^3Y^z
		\braket{\tout|a_-(\w\V x_z)\mS|\tin}+\hc.
	\label{soft1}
\end{align}
Here $\V x_z$ denotes the unit 3-vector pointing in the direction defined by $(z,\zb)$,
\begin{align}
	\V x_z = \left(\frac{\zb+z}{1+z\zb},\frac{i(\zb-z)}{1+z\zb},\frac{1-z\zb}{1+z\zb}\right).
\end{align}
Superrotation is a symmetry of the S-matrix, which implies that the superrotation charge
\begin{align}
	Q(Y) = Q_S(Y) + Q_H(Y),
\end{align}
is conserved in scattering process, that is, $\braket{\tout|[Q(Y),\mS]|\tin}=0$.
Equivalently,
\begin{align}
	\braket{\tout|[Q_S,\mS]|\tin} = -\braket{\tout|[Q_H,\mS]|\tin}.
\end{align}
Using \eqref{hard1} and \eqref{soft1}, this can be written as
\begin{align}
	&\frac{i}{2\pi\kappa}\lim_{\w\to 0}(1+\w\p_\w)\int d^2zD_z^3Y^z
		\braket{\tout|a_-(\w\V x_z)\mS|\tin}+\hc
	\nonumber \\&\qquad=
		i\sum_i \left(
			Y^z(z_i)\p_{z_i}-\frac{E_i}{2}D_zY^z(z_i)\p_{E_i}+\zz
		\right)\braket{\tout|\mS|\tin}.
	\label{conservation1}
\end{align}
For the vector field $Y^z$, let us choose
\begin{align}
	Y = \frac{(z-w)^2}{(\zb-\wb)}\p_z,
\end{align}
which satisfies $D_z^3Y^z = 4\pi\delta^{(2)}(z-w)$, as well as
\begin{align}
	D_z Y^z(z_i) &= \frac{2(w-z_i)(1+w \zb_i)}{(\wb-\zb_i)(1+z_i\zb_i)}.
	\label{DY}
\end{align}
Then \eqref{conservation1} becomes
\begin{align}
	&\lim_{\w\to 0}(1+\w\p_\w)
		\braket{\tout|a_-(\w\V x_w)\mS|\tin}
	\nonumber \\&\qquad=
		\frac{\kappa}{2}\sum_i \left(
			-\frac{(w-z)^2}{(\wb-\zb)}\p_{z_i}
			-\frac{E_i(w-z_i)(1+w \zb_i)}{(\wb-\zb_i)(1+z_i\zb_i)}\p_{E_i}
		\right)\braket{\tout|\mS|\tin}.
	\label{conservation2}
\end{align}
We now show that the factor in the parentheses on the RHS is the subleading soft factor,
\begin{align}
	S_g^{(1)-} = -i\frac{\kappa}{2}\sum_i\eta_i\frac{p_i^\mu k_\lambda J_i^{\lambda \nu}}{p_i\cdot k}\ep^-_\mn(\V k).
		\label{subleading_gravity}
\end{align}
Since all hard particles are assumed to be scalars, the angular momentum consists of only the orbital part,
\begin{align}
	\eta_i (J_i)_\mn = \eta_i (L_i)_\mn \equiv -i\left(p_{i\mu} \frac{\p}{\p p_i^\nu} - p_{i\nu} \frac{\p}{\p p_i^\mu}\right),
\end{align}
where $\eta_i=+1$ ($-1$) if $i$-th particle is outgoing (incoming).
Our definition of $J_\mn$ is different from that of \cite{Kapec:2014opa} by a sign, see footnote \ref{eta_origin}.
Now let us parametrize
\begin{align}
	\begin{split}
		p^\mu &= \frac{E}{1+z\zb}\left(1+z\zb,\zb+z,i(\zb-z),1-z\zb\right),\\
		k^\mu &= \frac{\w_k}{1+z\zb}\left(1+w\wb,\wb+w,i(\wb-w),1-w\wb\right),\\
		\ep^{-\mu} &= \frac{1}{\sqrt 2}(w,1,i,-w),
	\end{split}
	\label{transf}
\end{align}
and $\ep^{-\mn} = \ep^{-\mu}\ep^{-\nu}$.
The quantities $(E,z,\zb)$ are related to $p^\mu$ by
\begin{align}
	E = \sqrt{(p^x)^2 + (p^y)^2+(p^z)^2},
	\qquad
	z = \frac{p^x + ip^y}{p^t + p^z},
	\qquad
	\zb = \frac{p^x - ip^y}{p^t + p^z}.
\end{align}
Let us write \eqref{subleading_gravity} out as
\begin{align}
	S_g^{(1)-} &=
		-(p\cdot \ep^-)
		\left(\ep^{-\mu} \frac{\p}{\p p^\mu} - \frac{(p\cdot \ep^-)}{(p\cdot k)}k^\mu \frac{\p}{\p p^\mu}\right).
	\label{subleading}
\end{align}
One can show using \eqref{transf} that
\begin{align}
	p\cdot \ep^- &= \frac{-\sqrt 2 E(w-z)}{(1+z\zb)},
	\label{pe}
	\\
	p\cdot k &= -\frac{2\w_k E(w-z)(\wb-\zb)}{(1+w\wb)(1+z\zb)}.
	\label{pk}
\end{align}
Now, noting that
\begin{align}
	\frac{\p}{\p p^\mu} &= \frac{\p E}{\p p^\mu}\p_E + \frac{\p z}{\p p^\mu}\p_z + \frac{\p\zb}{\p p^\mu}\p_\zb,
\end{align}
we obtain
\begin{align}
	\frac{\p}{\p p^t} &= -\frac{z(1+z\zb)}{2E}\p_z - \frac{\zb(1+z\zb)}{2E}\p_\zb,
	\\
	\frac{\p}{\p p^x} &=\frac{(\zb+z)}{(1+z\zb)}\p_E + \frac{(1+z\zb)}{2E}\p_z + \frac{(1+z\zb)}{2E}\p_\zb,
	\\
	\frac{\p}{\p p^y} &=\frac{i(\zb-z)}{(1+z\zb)}\p_E + \frac{i(1+z\zb)}{2E}\p_z - \frac{i(1+z\zb)}{2E}\p_\zb,
	\\
	\frac{\p}{\p p^z} &=\frac{(1-z\zb)}{(1+z\zb)}\p_E - \frac{z(1+z\zb)}{2E}\p_z - \frac{\zb(1+z\zb)}{2E}\p_\zb.
\end{align}
Using these with \eqref{pe} and \eqref{pk}, we may write \eqref{subleading} as
\begin{align}
	S_g^{(1)-} = \frac{\kappa}{2}\sum_i\left(-\frac{E_i(w-z_i)(1+w\zb_i)}{(\wb-\zb_i)(1+z_i\zb_i)}\p_{E_i}-\frac{(w-z_i)^2}{(\wb-\zb_i)}\p_{z_i}\right).
	\label{subleading_gravity2}
\end{align}
This is exactly the expression appearing on the RHS of \eqref{conservation2}, which was to be shown.

\subsubsection{Spin correction}\label{app:gravspin}

Now suppose the hard particles have non-zero spin.
The angular momentum $J^\mn$ appearing in the subleading soft factor now contains the spin contribution $S^\mn$,
\begin{align}
	J^\mn = L^\mn + S^\mn.
\end{align}
Let us define helicity in terms of the Pauli-Lubansky pseudovector,
\begin{align}
	h p_\mu &= -\frac{1}{2}\ep_\mnrs J^{\nu\rho}p^\sigma
		=-\frac{1}{2}\ep_\mnrs S^{\nu\rho}p^\sigma,
		\label{PL}
\end{align}
where $\ep_\mnrs$ is the Levi-Civita tensor with $\ep_{0123}=1$, and in the second equation the orbital
	part drops out due to antisymmetry in $p$.
In this basis, the spin angular momentum has components \cite{HILGEVOORD19651002}
\begin{align}
	S_\mn = \frac{h}{E}
		\begin{pmatrix}
			0&0&0&0 \\
			0&0&p^z&-p^y\\
			0&-p^z&0&p^x\\
			0&p^y&-p^x&0
		\end{pmatrix}
	= \frac{h}{1+z\zb}
		\begin{pmatrix}
			0&0&0&0 \\
			0&0&1-z\zb&-i(\zb-z)\\
			0&-(1-z\zb)&0&\zb+z\\
			0&i(\zb-z)&-(\zb+z)&0
		\end{pmatrix},
	\label{spin}
\end{align}
where in the second equation we used the parametrization \eqref{transf}.

For particles with spin, the action of the hard charge is \cite{Kapec:2014opa}
\begin{align}
	\braket{\tout|[Q_H,\mS]|\tin} &=
		i\sum_i \left(
			\mL_Y-\frac{E_i}{2}D_AY^A(z_i)\p_{E_i}
		\right)\braket{\tout|\mS|\tin},
	\label{hard3gravity}
\end{align}
where $\mL_Y$ is the Lie derivative on the 2-sphere with respect to $Y$,
\begin{align}
	\mL_Y = Y^A\p_A + \frac{i}{2}D_{A}Y_{B}S^{AB},\qquad A,B\in\{z,\zb\},
\end{align}
where $S_{AB}$ is the pullback of \eqref{spin} to the 2-sphere.
By coordinate transformation from $\hat x^\mu$ to $(z,\zb)$, one finds that
\begin{align}
	S_{\zb z} &= \frac{\p \hat x^\mu}{\p \zb}\frac{\p \hat x^\nu}{\p z} S_\mn = -\frac{2ih}{(1+z\zb)^2} = -ih\gzz,
\end{align}
and similarly $S_{z\zb} = ih\gzz$, $S_{zz}=S_{\zb\zb}=0$.
Thus,
\begin{align}
	\mL_Y = Y^z\p_\zb + \frac{h}{2}D_{z}Y^z + Y^\zb\p_\zb - \frac{h}{2}D_\zb Y^\zb,
\end{align}
and \eqref{hard3gravity} can be written as
\begin{align}
	\braket{\tout|[Q_H,\mS]|\tin} &=
		i\sum_i \bigg[
			Y^z(z_i)\p_{z_i}-\frac{E_i}{2}D_zY^z(z_i)\p_{E_i}+\frac{h_i}{2}D_zY^z(z_i)
			+(z\to \zb,h\to -h)
		\bigg]
		\nonumber\\&\qquad \times\braket{\tout|\mS|\tin}.
	\label{hard2gravity}
\end{align}
In the previous section, we saw that with the choice $Y = (z-w)^2(\zb-\wb)^{-1}\p_z$,
	the first two terms in the square brackets correspond to the subleading soft factor coming from the orbital angular momentum $L^\mn$.
The third term then must correspond to the factor coming from spin angular momentum $S^\mn$, which is
\begin{align}
	-i\sum_i \frac{p_i^\mu k_\lambda S_i^{\lambda\nu}}{p_i\cdot k}\ep^-_\mn.
\end{align}
We have already computed $p_i\cdot \ep^-$ and $p_i\cdot k$ in \eqref{pe} and \eqref{pk}.
Using \eqref{spin} and \eqref{transf}, one can directly show that
\begin{align}
	k^\lambda S_{\lambda \nu} \ep^{-\nu} = \frac{\sqrt 2 i h_i\w_k(w-z_i)(1+w\zb_i)}{(1+w\wb)(1+z_i\zb_i)}.
\end{align}
Combining the results, we obtain
\begin{align}
	-i \frac{p_i^\mu k_\lambda S_i^{\lambda\nu}}{p_i\cdot k}\ep^-_\mn
		= \frac{h_i(w-z_i)(1+w\zb_i)}{(\wb-\zb_i)(1+z_i\zb_i)}.
		\label{spin_gravity}
\end{align}
But from \eqref{DY} we observe that this can be written as
\begin{align}
	-i \frac{p_i^\mu k_\lambda S_i^{\lambda\nu}}{p_i\cdot k}\ep^-_\mn
		= \frac{h_i}{2}D_zY^z(z_i),
\end{align}
which is exactly the third term in the square brackets of \eqref{hard2gravity},
	showing that the formalism extends to particles with spin.

\subsection{QED}\label{app:qed}

The results of appendix \ref{app:gravspin} is directly relevant to QED since charged particles have spin.
The construction is very similar -- we will start with charged scalars and move on to spin corrections.

\subsubsection{Massless scalars} \label{app:qedscalar}

From the action of hard and soft charges for charged scalars, we have \cite{Lysov:2014csa}
\begin{align}
	\braket{\tout|[\mQ_H,\mS]|\tin} &=
		i\sum_iQ_i \left(
			\frac{1}{E_i}Y^z(z_i)\p_{z_i}-D_zY^z(z_i)\p_{E_i}+\zz
		\right)\braket{\tout|\mS|\tin},
	\label{hard1qed}
	\\
	\braket{\tout|[\mQ_S,\mS]|\tin} &=
		-\frac{i}{4\pi e}\lim_{\w\to 0}(1+\w\p_\w)\int d^2zD_z^2Y^z\frac{\sqrt 2}{1+z\zb}
		\braket{\tout|a_-(\w\V x_z)\mS|\tin}+\hc.
	\label{soft1qed}
\end{align}
Now let us choose the vector field
\begin{align}
	Y=\frac{(z-w)(1+z\zb)}{(\zb-\wb)}\p_z,
\end{align}
which satisfies $D_z^2Y^z=2\pi(1+z\zb)\delta^{(2)}(z-w)$ and
\begin{align}
	D_zY^z = -\frac{(1+w\zb)}{(\wb-\zb)}.
\end{align}
This leads the charge conservation $\braket{\tout|[\mQ_S,\mS]|\tin}=-\braket{\tout|[\mQ_H,\mS]|\tin}$ to be written as
\begin{align}
	\lim_{\w\to 0}(1+\w\p_\w)&
		\braket{\tout|a_-(\w\V x_z)\mS|\tin}
		\nonumber\\&=
		\frac{e}{\sqrt 2}\sum_iQ_i \left(
			\frac{(w-z_i)(1+z_i\zb_i)}{E_i(\wb-\zb_i)}\p_{z_i}+\frac{(1+w\zb_i)}{(\wb-\zb_i)}\p_{E_i}
		\right)\braket{\tout|\mS|\tin}.
		\label{conservation_qed}
\end{align}
We now show that the factor on the RHS is the subleading soft factor for scalars,
\begin{align}
	S_e^{(1)-} = -ie\sum_i\eta_i Q_i \frac{k_\lambda L_i^{\lambda\nu}}{p_i\cdot k}\ep^-_\nu.
\end{align}
Comparing this to gravity \eqref{subleading_gravity}, we observe that this is just
	 $S_g^{(1)-}$ times $2eQ_i[\kappa(p_i\cdot \ep^-)]^{-1}$.
From \eqref{pe} and \eqref{subleading_gravity2}, we thus obtain
\begin{align}
	S_e^{(1)-} &=
		\frac{e}{\sqrt 2}\sum_iQ_i\left(
			\frac{(w-z_i)(1+z_i\zb_i)}{E_i(\wb-\zb_i)}\p_{z_i}
			+\frac{(1+w\zb_i)}{(\wb-\zb_i)}\p_{E_i}
		\right),
	\label{qed_scalar_subleading}
\end{align}
which exactly agrees with the RHS of \eqref{conservation_qed}.

\subsubsection{Spin correction}

In the presence of spin, the subleading factor \eqref{qed_scalar_subleading} gains an additional spin contribution, which can be obtained
	using the results of appendix \ref{app:gravspin}.
We shall show that this contribution is exactly the spin corrected term in the action of hard charge obtained by replacing
	$Y^z\p_z$ with $\mL_Y$ in \eqref{hard1qed}.

Just as the orbital angular momentum piece of $S_e^{(1)-}$, multiplying $eQ_i(p_i\cdot \ep^-)^{-1}$ to \eqref{spin_gravity} should
	give us the spin angular momentum part, i.e.
\begin{align}
	-ieQ_i \frac{ k_\lambda S_i^{\lambda\nu}}{p_i\cdot k}\ep^-_\mu
		&=-\frac{eQ_ih_i}{\sqrt 2E_i}\frac{(1+w\zb_i)}{(\wb-\zb_i)}.
	\label{spinqed}
\end{align}
Thus the full subleading soft factor becomes
\begin{align}
	S_e^{(1)-} &=
		\frac{e}{\sqrt 2}\sum_iQ_i
		\left(
			\frac{(1+w\zb_i)}{(\wb-\zb_i)}\p_{E_i}
			+\frac{(w-z_i)(1+z_i\zb_i)}{E_i(\wb-\zb_i)}\p_{z_i}
			-h_i\frac{(1+w\zb_i)}{E_i(\wb-\zb_i)}
		\right).
\end{align}
The spin-corrected action of hard charge is obtained by replacing $Y^z\p_z$ with $\mL_Y$ in \eqref{hard1qed},
\begin{align}
	\braket{\tout|[\mQ_H,\mS]|\tin} &=
		i\sum_iQ_i \left(
			\frac{1}{E_i}\mL_Y-D_zY^z(z_i)\p_{E_i}+(z\to \zb)
		\right)\braket{\tout|\mS|\tin}
	\\ &=
		i\sum_iQ_i \left(
			\frac{1}{E_i}Y^z(z_i)\p_{z_i}+\frac{h}{E_i}D_zY^z(z_i)-D_zY^z(z_i)\p_{E_i}+(z\to \zb,h\to -h)
		\right)
		\nonumber\\&\qquad\times\braket{\tout|\mS|\tin}.
\end{align}
Accordingly, the charge conservation \eqref{conservation_qed} becomes
\begin{align}
	&\lim_{\w\to 0}(1+\w\p_\w)
		\braket{\tout|a_-(\w\V x_z)\mS|\tin}
		\nonumber\\&=
		\frac{e}{\sqrt 2}\sum_iQ_i \left(
			\frac{(w-z_i)(1+z_i\zb_i)}{E_i(\wb-\zb_i)}\p_{z_i}
			+\frac{(1+w\zb_i)}{(\wb-\zb_i)}\p_{E_i}
			-h_i\frac{(1+w\zb_i)}{E_i(\wb-\zb_i)}
		\right)\braket{\tout|\mS|\tin}.
\end{align}
The new term is exactly the spin contribution \eqref{spinqed} to the subleading soft factor $S_e^{(1)-}$.

\bibliographystyle{jhep} 
\bibliography{subleading}

\end{document}